\documentclass[prb,twocolumn,aps,showpacs,numerical]{revtex4}
\usepackage{graphicx,simplewick}
\usepackage{hyperref}
\usepackage{amsmath}
 \usepackage[utf8]{inputenc}
 \usepackage[T1]{fontenc}
\usepackage{amssymb}
\usepackage{natbib}
\usepackage{amstext}
\usepackage{amsfonts}
\usepackage{epstopdf}
\DeclareMathOperator{\sgn}{sgn}
\newcommand{\hphi}{\hat{\phi}}

\begin{document}

\title{Strongly correlated dynamics in multichannel quantum RC circuits}								
\author{Prasenjit Dutt$^1$, Thomas L. Schmidt$^{2,1}$, Christophe Mora$^{3}$, and Karyn Le Hur$^{4,1}$}
\affiliation{$^1$ Department of Physics, Yale University, New Haven, CT 06520, USA}
\affiliation{$^2$ Departement Physik, Universt\"at Basel, Klingelbergstrasse 82, 4056 Basel, Switzerland}
\affiliation{$^3$ Laboratoire Pierre Aigrain, \'Ecole Normale Sup\'erieure, Universit\'e Denis Diderot, CNRS; 24 rue Lhomond, 75005 Paris, France}
\affiliation{$^4$ CPHT, \' Ecole Polytechnique, CNRS, Palaiseau 91128 C\' edex, France}

\date{\today}

\begin{abstract}
We examine dissipation effects in a multichannel quantum RC circuit, comprising a cavity or single-electron box capacitively coupled to a gate and connected to a reservoir lead via several conducting channels. Depending on the engineering details of the quantum RC circuit, the number of channels contributing to transport vary, as do the form of the interchannel couplings. For low-frequency AC transport, the charge-relaxation resistance ($R_{q}$) is a nontrivial function of the parameters of the system. However, in the vicinity of the charge degeneracy points and for weak tunneling, we find as a result of cross-mode mixing or channel asymmetry that $R_q$ becomes universal for a metallic cavity at low temperatures, and equals the unit of quantum resistance. To prove this universality we map the system to an effective one-channel Kondo model, and construct an analogy with the Coulomb gas. Next, we probe the opposite regime of near-perfect transmission using a bosonization approach. Focussing on the two-channel case, we study the effect of backscattering at the lead-dot interface, more specifically, the role of an asymmetry in the backscattering amplitudes, and make a connection with the weak tunneling regime near the charge degeneracy points.
\end{abstract}


\maketitle

\section{Introduction}

The manipulation of mesoscopic systems to engineer quantum circuits has immense potential for future applications. These unique systems exhibit a spectrum of novel phenomena which necessitates a  better understanding of their dynamics. Technological advances have provided the means to couple these systems to capacitive gates, thereby enabling detailed exploration of electronic transport at the nanoscale.

\begin{figure}[t]
	\centering
		\includegraphics[width=0.95\columnwidth]{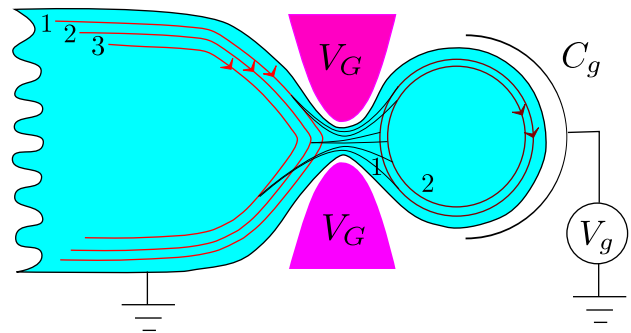}
	\caption{Schematic of a multichannel quantum RC circuit where ${\cal M}=3$ and ${\cal N}=2$, where ${\cal M}$ denotes the number of channels in the lead and ${\cal N}$ denotes the number of channels in the dot.  All the possible lead-dot couplings are depicted by black lines. The AC drive is capacitively coupled to the single-electron box. The geometric capacitance $C_g$, is to be distinguished from the mesoscopic capacitance $C_0$ of the quantum RC circuit.}
	\label{fig:RC_schematic}
\end{figure}

In particular, the phenomenon of Coulomb blockade offers an excellent tool for the observation of interaction effects at the nanoscale \cite{Wahlgren,Averin,Nazarov} and has become one of the cornerstones of modern condensed-matter physics. One of the simplest mesoscopic systems exhibiting Coulomb blockade is the single-electron box (quantum dot) and almost two decades have elapsed since the first experimental evidence of macroscopic charge quantization\cite{Esteve}. Quantum coherence and interaction effects drastically affect the properties of these systems. In fact, dissipation and resistance in single-electron boxes have recently sparked a growing attention in several parametric regimes both theoretically \cite{Thomas2,Thomas,Buttiker1,Imry2,Ioselevich,Martin,ChristKaryn,Martin2,Filippone,Etzioni1,Etzioni2,Governale,Kashuba} and experimentally \cite{Gabelli,Feve,Review,Ihn}. Specifically, the linear charge response to a gate voltage oscillation for single-electron devices has gained considerable attention. 

In this paper, we investigate a multichannel version of the quantum Resistance-Capacitance (RC) circuit. A possible realization of this system involves a Coulomb-blockaded quantum dot coupled via a quantum point contact to a two-dimensional electron gas, which is in turn capacitively connected to a back gate (see Fig. \ref{fig:RC_schematic}). By tuning the opening of the quantum point contact \cite{Wharam,Wees,ButtikerQPC,Imry,Beenakker,Kontos} through an auxiliary gate voltage, one can suitably control the number of conduction channels transmitted through the cavity. Here, we examine the low frequency behavior of the charge relaxation resistance in the cavity, in the limit of low temperatures such that quantum coherence is preserved \cite{Thomas2,Thomas}. For a small box, the role of coherence in charge quantization for a box consisting of noninteracting electrons has been studied \cite{Nigg_2}. In this work we investigate the role of Coulomb interactions in the charge quantization of the multichannel quantum RC circuit in the opposite regime of a large dot.
For a single conducting channel (mode), the relaxation of the charge on the dot when subject to an AC drive voltage has been studied by B\"uttiker {\it et. al.} \cite{Thomas2,Thomas,Buttiker1}. The authors were the first to predict the universality of the charge relaxation resistance $(R_q)$ in the context of a small (coherent) cavity, where the Coulomb interactions have been treated within the Hartree-Fock approximation. The quantum RC circuit has been realized experimentally in a two-dimensional electron gas, and the predicted charge relaxation resistance $R_q=h/(2e^2)$ has been confirmed \cite{Gabelli,Feve,Ihn}. 

Recently, the robustness of this value of quantized resistance in the presence of interactions in the cavity has been rigorously proved \cite{ChristKaryn,Martin}. It has been shown that there is a mesoscopic crossover of the charge relaxation resistance from $R_q=h/(2e^2)$ to $R_q=h/e^2$ as the size of the quantum dot is increased \cite{ChristKaryn}. The value $R_q=h/e^2$ for a large (metallic) cavity in the vicinity of a charge degeneracy point has been obtained via an analogy with the Kondo model \cite{Berman,Lehnert} and is a consequence of the emergent Fermi liquid ground state \cite{Nozieres,Filippone,Filippone2,Garst}. In this regime, an electron entering into the cavity is disentangled from an electron escaping the cavity. This unit of resistance can be viewed as two Sharvin-Imry contact resistances in series. The charge relaxation resistance effectively captures interaction effects in a variety of exotic systems, such as fractional quantum Hall edge states \cite{Martin,ChristKaryn,Grenier}, the Anderson impurity model \cite{Martin,Filippone} and topological insulator edge states \cite{IonKaryn}, all of which can be used to construct a quantum RC setup. The dynamical charge response in the case of a Majorana Coulomb box has also been analyzed \cite{Golub,Mora_LeHur_2}.

For a metallic cavity increasing the number of conducting channels through the constriction causes the transport away from the charge degeneracy points to deviate from $h/e^2$, and the value of $R_q$ is very sensitive to the engineering details of the system. The central message of this paper is that close to the charge degeneracy points however, when the tunneling couplings are weak, the charge relaxation resistance becomes
universal and is in fact independent of the number of channels.  To prove this result we project out states apart from those in the immediate vicinity of the charge-degeneracy point and reformulate the problem in terms of a pseudospin \cite{Matveev,Aleiner,Zarand,LeHur,Lebanon}, and show via an analogy with the Coulomb gas \cite{Anderson1,Anderson2} that the system can be mapped to the single-channel anisotropic Kondo model. 

The dynamics of the system in the opposite regime of perfect or near perfect transmission, {\it i.e.} strong tunneling, can be studied using bosonization techniques. Adopting a generalization of well-known bosonization methods  \cite{Matveev2,Glazman,Aleiner,LeHur,Flensberg,ChristKaryn} to ${\cal N}$ channels, we show that $R_q$ is non-universal. For reflectionless channels with equal tunneling amplitudes $R_q=h/\left({\cal N}e^2\right)$. We focus on the special case ${\cal N}=2$, and study the effect of backscattering, in particular asymmetry in the reflection amplitudes, on the value of $R_q$. For the Coulomb blockaded dot, the charge on the dot is pinned. We analyze the effect of charge fluctuations in the dot on the value of $R_q$ for weak backscattering at the dot-lead interface. We show that a second order calculation in the backscattering amplitudes, valid at high temperatures and frequencies, does not correct this value of $R_q$.  However, based on renormalization group arguments we conclude that the value $h/e^2$ of $R_q$ reemerges at low-energies, where the system flows to strong backscattering, in agreement with the weak tunneling analysis. More precisely, the system can be mapped onto an anisotropic two-channel Kondo model at the Emery-Kivelson line \cite{Emery-Kivelson}. An asymmetry in the backscattering amplitudes near the charge-degeneracy point causes the system to flow to a one-channel Kondo model at low-energies, since the channel with a stronger (bare) backscattering amplitude is eventually perfectly reflected and pinched off \cite{LeHur}.

In a different context, the variation of $R_q$ of a Coulomb box for a Landau-Zener sweep of the gate voltage has been studied by mapping the system to a dissipative particle confined to a ring\cite{Etzioni1,Etzioni2}. For this out-of-equilibrium situation it has been shown that when a quantum of magnetic flux is passed through the ring, the average of the relaxation resistance $\bar{R}_q = h/e^2$.  

The remainder of the paper is organized as follows. In Sec. II, we present our results away from the charge degeneracy points via a perturbative expansion in the tunneling Hamiltonian following Ref. \onlinecite{ChristKaryn}. In Sec. III, we discuss the underlying Kondo physics in the vicinity of a charge degeneracy point and formulate an analogy with the Coulomb gas\cite{Anderson1,Anderson2}. In Sec. IV, we briefly address the situation at and close to perfect  transmission. Appendices are devoted to technical details and mathematical derivations.

\section{Weak tunneling analysis away from the charge degeneracy points}\label{weak_tunneling}

The Hamiltonian of the system is given by
\begin{align}\label{eq:hamiltonian_RC} H&=\sum_{k\alpha} \epsilon_{k} \, d^\dagger_{k\alpha} d_{k\alpha}+\sum_{p\beta} \varepsilon_{p} \, c^\dagger_{p\beta} c_{p\beta} + E_c (\hat{N}-N_0)^2\notag\\ 
&\qquad+\sum_{kp\alpha\beta}t_{\alpha\beta} \left( d^\dagger_{k\alpha} c_{p\beta} + c^\dagger_{p\beta} d_{k\alpha} \right).
\end{align}
 Here, the subscripts $\alpha(=1,\ldots,{\cal N})$ and $\beta(=1,\ldots,{\cal M})$ denote the channel index in the dot and lead respectively. At a general level, the channel (mode) index can also account for spin degrees of freedom keeping in mind that tunneling between different spin projections is inadmissible. Hereafter, we consider the limit of large metallic cavity with a dense spectrum of energy levels. The single electron eigenfunctions of the two-dimensional system and the quantum dot are labeled by the index $p$ and $k$ respectively. The charging energy of the dot is expressed in terms of the charge operator 
 \begin{equation}
 \hat{Q}=e\hat{N}=e\sum_{k\alpha}d^\dagger_{k\alpha} d_{k\alpha}
 \end{equation}
  on the dot and $N_0=C_g V_{g}/e$ is imposed by the gate voltage (see Ref. \onlinecite{ChristKaryn}). We allow for a variety of inter-channel and intra-channel tunneling amplitudes ($t_{\alpha\beta}$), depending on how the system is engineered. Note that in the following, we work at very low temperatures $(T\rightarrow 0)$ in order to preserve the quantum coherence. 

In this Section, we derive the dependence of the relaxation resistance $R_q$ on the number of channels $({\cal M, N})$ in the limit of weak tunneling amplitudes away from the charge degeneracy points of the dot, as given by Eq.\ \eqref{eq:hamiltonian_RC}. We assume that the charging energy $E_c$ is
the most dominant energy scale, which implies the Coulomb blockade limit \cite{Nazarov}. The computation is a generalization of the scheme presented in Ref.\ [\onlinecite{ChristKaryn}]. In this regime it is useful to group the terms in the Hamiltonian as follows,
\begin{subequations}\label{weak_hamil}
\begin{align}
\label{h_wt} H&=H'+H_{\text{T}}, \\[1mm]
\label{h_ni} H'&=\sum_{k\alpha} \epsilon_{k} \, d^\dagger_{k\alpha} d_{k\alpha}+\sum_{p\alpha} \varepsilon_{p} \, c^\dagger_{p\beta} c_{p\beta} + E_c (\hat{N}-N_0)^2, \\[1mm]
\label{h_pert} H_{\text{T}} &= \sum_{kp\alpha\beta}t_{\alpha\beta} \left( d^\dagger_{k\alpha} c_{p\beta} + c^\dagger_{p\beta} d_{k\alpha} \right).
\end{align}
and treat the tunneling term $H_\text{T}$ as a perturbation. The channels of the dot are denoted by $\alpha(=1,\ldots,{\cal N})$, whereas for the lead it is given by $\beta(=1,\ldots,{\cal M})$.
\end{subequations}
We study the AC response of the circuit using linear response theory. The gate voltage can be separated into its AC and DC components 
\begin{align}
 V_{g}(t)=V_{g0} + V_{g1}(t).
\end{align}
In the presence of a small time-dependent perturbation of the gate
voltage, the charge on the dot $Q = e \langle \hat{N} \rangle$ obeys,
\begin{equation}\label{linear}
Q (\omega) = e^2\tilde{K} (\omega) V_g (\omega),
\end{equation}
where the retarded response function, following standard linear response theory
\begin{equation}
\tilde{K} ( t - t') = i \theta(t-t')\left\langle [\hat{N} (t), \hat{N} (t')]\right\rangle
\end{equation}
measures the charge fluctuation induced by the AC component of the gate voltage.
In the absence of electron tunneling, the cavity charge in the ground state is $\langle\hat{N}\rangle = N^{\ast}$ and does not fluctuate, hence $\tilde{K} = 0$. We are at freedom to set $N^{\ast}=0$. Assuming weak tunneling, the charge fluctuations on the cavity are determined using perturbation theory in $H_\text{T}$. At $T=0$ the retarded and the time-ordered Green's function have a simple relation in frequency domain, so we instead compute
\begin{align}
\label{K}
K(t-t')=i\left\langle T_{t}[\hat{N} (t)\hat{N} (t') ]\right\rangle.
\end{align}
We use the perturbative expansion in the interaction picture
\begin{align}\label{correl}
K(t)&=i\langle T_t [\hat{N}_{I}(t) \hat{N}_{I}(0)]\rangle \notag\\
&= i\frac{\langle \phi_{GS}|
T_t [\hat{N}_{I}(t) \hat{N}_{I}(0) U_{I}(\infty,-\infty) ]|\phi_{GS}\rangle}
{\langle \phi_{GS} |
T_t [ U_{I}(\infty,-\infty) ] | \phi_{GS}\rangle}.
\end{align}
Since the unperturbed Hamiltonian $H'$ is not quadratic, Wick's theorem is not applicable. We therefore expand the evolution operator
\begin{align}
 U_{I}(\infty,-\infty) &= \sum_{n=0}^{\infty} \left ( -i \right)^n
\int_{-\infty}^{\infty} dt_1 \int_{-\infty}^{t_1} dt_2 \ldots\notag\\
&\ldots  \int_{-\infty}^{t_{n-1}} dt_n H_\text{T}(t_1) H_\text{T} (t_2) \ldots H_\text{T}(t_n),
\end{align}
in powers of $ H_\text{T}(t) = e^{i H' t} H_\text{T}(0)  e^{-i H' t}$. In the ground state $\hat{N}|\phi_{GS}\rangle$=0. Thus, the zeroth and first order contribution to $K(\omega)$ vanish. The leading order contribution arises at second order. Using equations of motion the time-evolved electron annihilation operators can be expressed as
\begin{align}
\label{eq:c(t)}
 c_{p\beta}(t)&=e^{-i\varepsilon_{p}t}c_{p\beta},\\
\label{eq:d(t)}
 d_{k\alpha}(t)&=e^{-i(\epsilon_{k}+E_{C}(2\hat{N}-2N_{0}+1))t}d_{k\alpha}.
\end{align}

The charge relaxation resistance at low frequency is given by the expression:
\begin{equation}
\label{charge}
\frac{Q(\omega)}{V_g(\omega)} = C_0(1+ i\omega C_0 R_q) + {\cal O}(\omega^2),
\end{equation}
the details of which are given in Appendix \ref{app:A}. The mesoscopic capacitance $C_0$, which is distinct from the geometrical capacitance $C_g$, in the weak-tunneling limit gives
\begin{equation}
\label{C0}
C_0 = \frac{\nu_0 \nu_1 C_g}{1/4-N_0^2}\sum_{\alpha\beta} t_{\alpha\beta}^2,
\end{equation}
where $\nu_0$ and $\nu_1$ represent the density of states in the lead and in the metallic cavity, respectively. Here we have retained terms to second-order in $H_\text{T}$. A fourth-order computation $H_\text{T}$ allows us to extract the leading-order dissipative (purely imaginary) contribution to the function $K(\omega)$ in Eq. (\ref{linear}). Thus, we obtain
\begin{align}
\label{Rqb}
 R_{q}=\frac{2\pi\hbar}{e^{2}}\left[\frac{ \sum_{\genfrac{}{}{0pt}
{}{\alpha_1\beta_1}{\alpha_2\beta_2}}t_{\alpha_1\beta_1}t_{\alpha_2\beta_1}t_{\alpha_2\beta_2} t_{\alpha_1\beta_2}}{\left(\sum_{\alpha\beta}t_{\alpha\beta}^{2}\right)^{2}}\right].
\end{align}
This illustrates that in general, away from the charge degeneracy points, {\it i.e.}, for $N_0\neq 1/2$, the charge relaxation resistance $R_q$ is not universal as it depends on the engineering details of the system. For the case of diagonal couplings $t_{\alpha\beta}=t_{\alpha}\delta_{\alpha\beta}$ (${\cal N}={\cal M}$), $R_{q}$ reduces to the non-universal result 
 \begin{align}
 \label{Rq}
 R_{q}=\frac{h}{e^{2}}\left[\frac{\sum_{\alpha}t_{\alpha}^{4}}{\left(\sum_{\alpha}t_{\alpha}^{2}\right)^{2}}\right].
\end{align}
It should be noted that in the case of (almost) isotropic diagonal couplings, {\it i.e.}, $t_{\alpha}=t$, the quantum RC circuit
becomes dissipationless in the limit ${\cal N}\rightarrow \infty$ as $R_q=h/(e^2{\cal N})$. This value demonstrates the violation of Kirchhoff's law, which in contrast predicts that $R_q$ would be inversely proportional to the sum of the transmission probabilities  \cite{Feve,Gabelli}, and is a direct consequence of the quantum (phase) coherence. 

We remark that for the special case when the transmission prabability of a single channel dominates ,{\it i.e.} $t_1\gg\{t_2,\ldots,t_{\mathcal N}\}$ one recovers the value $R_q=h/e^2$.

\section{Kondo model and Coulomb gas near the charge degeneracy points}
 
The weak-tunneling analysis of Section II leads to the surprising conclusion that for the case of large inter-channel mixing $R_q=h/e^2$, {\it i.e.} a reemergence of the unit of resistance. This follows directly from Eq.\ \eqref{Rqb} if we take the amplitudes of the tunneling matrix to be approximately of equal strength, {\it i.e.} $t_{\alpha\beta}\approx t$. It is possible to reformulate the quasiparticles of the dot and lead (individually) in terms of the totally symmetric combination of the modes and additionally construct ${\cal N}-1$ and ${\cal M}-1$ mutually orthonormal modes respectively. In this new formulation $t_{\alpha\beta}= t$ is the condition for perfect reflection of the additional ${\cal N}-1$ and ${\cal M}-1$  modes respectively, and effectively only a single channel (the totally symmetric mode) is transmitted. The value $R_q=h/e^2$ is a consequence of this emergent one-channel quantum RC circuit.

Due to the strong Coulomb blockade in the vicinity of a charge degeneracy point $N_0=n+1/2$ ($n\in {\mathbb Z}$), we can project out all charge states other than those corresponding to $n$ and $n+1$, which then mimic a spin-1/2 particle \cite{Matveev}. We show that this gives rise to an emergent one-channel Kondo model and also arrive at this result by mapping the system to the Coulomb gas \cite{Anderson1,Anderson2}. 

\subsection{Mapping on a Kondo model}

For the Coulomb blockaded dot near the charge degeneracy points $N_0=n+1/2$, the effective Hamiltonian of the system can be written as \cite{Matveev}
\begin{align}
\label{Kondo}
 H=&\sum_{k\alpha}\epsilon_{k} d^{\dag}_{k\alpha} d_{k\alpha}+\sum_{p\beta}\varepsilon_{p} c^{\dag}_{p\beta} c_{p\beta}-h_z S_z\notag\\
 &+\sum_{kp\alpha\beta}t_{\alpha\beta}\left(d^{\dag}_{k\alpha}c_{p\beta} S^{-}+c^{\dag}_{p\beta}d_{k\alpha} S^{+}\right).
 \end{align}
The effective magnetic field $h_z=e \left[N_0-\left(n+1/2\right)\right]/C_g$ denotes the deviation from the charge degeneracy points\cite{Matveev} and is assumed to be small compared to $E_c=e^2/2C_g$. Let us consider purely diagonal couplings, {\it i.e.}, for $t_{\alpha\beta}=t_{\alpha}\delta_{\alpha\beta}$ (${\cal N}={\cal M}$). It is straightforward to see from Eq.\ \eqref{Kondo} that one recovers the anisotropic ${\cal N}$-channel Kondo model
\begin{align}
  H=&\sum_{k\alpha}\epsilon_{k} d_{k\alpha} d_{k\alpha}+\sum_{p\beta}\varepsilon_{p} c^{\dag}_{p\beta} c_{p\beta}-h_z S_z\notag\\
  &+\sum_{kp\alpha}t_{\alpha}\left(d^{\dag}_{k\alpha}c_{p\alpha} S^{-}+c^{\dag}_{p\alpha}d_{k\alpha} S^{+}\right).
\end{align}
\begin{figure}[t]
	\includegraphics[width=0.95\columnwidth]{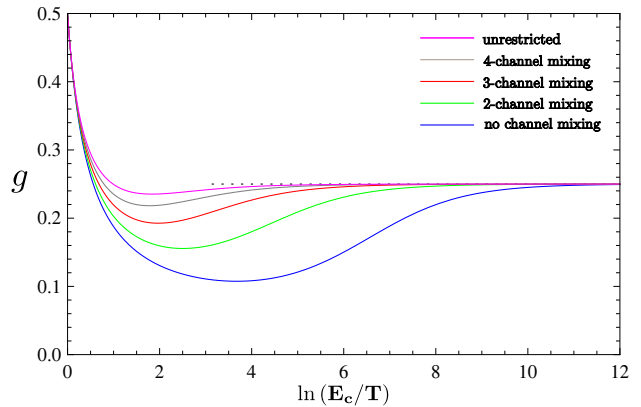}
	\caption{Renormalization of the DC conductance for a multichannel quantum RC circuit (${\cal M}={\cal N}=5$) with decreasing temperature, at the charge-degeneracy point. 
We consider randomly generated couplings (unrestricted) and eliminate progressively the interchannel-couplings, starting with the farthest channels(4-channel mixing) till all the inter-channel couplings are removed(no channel mixing). The random couplings are normalized according to the degree of channel mixing, such that $g(0)=0.5$. The dashed curve corresponds to the unitary limit of the Kondo model. } 
	\label{fig:eff_one_channel}
\end{figure}
Introducing a pseudospin index ($\sigma=\uparrow, \downarrow$) which denotes the position of a particle in the system, reservoir lead ($\uparrow$) versus cavity($\downarrow$). Furthermore, defining $d_{k\alpha}=a_{k\uparrow\alpha}$ and $c_{p\alpha}=a_{k\downarrow\alpha}$ we can rewrite the Hamiltonian of the system as
\begin{align}
 H_{\text{diag}}=&\sum_{k\alpha\sigma}E_{k\sigma} a^{\dag}_{k\sigma\alpha} a_{k\sigma\alpha}-h_z S_z\notag\\
 &+\sum_{kp\alpha}t_{\alpha}\left(a_{k\uparrow\alpha}^{\dag} a_{k\downarrow\alpha} S^{-}+ a_{k\downarrow\alpha}^{\dag}  a_{k\uparrow\alpha} S^{+}\right),
 \label{eq:matveev_N_channel}
\end{align}
which is precisely a ${\cal N}$-channel Kondo model with $J_{\perp\alpha}=t_{\alpha}$ and $J_{||\alpha}=0$ \cite{NozieresBlandin,Tsvelik,Andrei,Coleman,Jerez,Affleck}. Here, we assume that 
$E_{k\uparrow}=\epsilon_k=E_{k\downarrow}$. A small deviation from the resonance condition plays the role of a magnetic field, leading to a Zeeman term in the Kondo model. 

This equivalence with the ${\cal N}$-channel Kondo model leads to the value $R_q=h/({\cal N}e^2)$ and corroborates the prediction of Eq.\ \eqref{Rq}  in the weak-tunneling limit away from the charge-degeneracy points. The two-channel situation has been studied in detail by two of us \cite{Mora_LeHur_2}. 

As emphasized in Ref. \onlinecite{Zarand}, the general situation in the single-electron box is fairly intricate. In particular, cross-mode (channel) tunneling is a relevant perturbation which leads the system ultimately to a {\bf one-channel} Kondo fixed point in the case of spin-polarized electrons. This follows from the low-temperature  effective Hamiltonian given by Eq.\ \eqref{Kondo}, which predicts the emergence of a unique effective tunneling mode in the lead and the electron box. The other modes are perfectly back-scattered when the inter and intra-channel couplings are identical. Deviations from this condition causes partial transmission of these other ${\cal M}-1$ and ${\cal N}-1$ channels respectively, and leads to corrections of $R_q$. However, renormalization group analyses show that these perturbations are irrelevant, and for low energies the value of $R_q$ is  stable\cite{Zarand}. The energy scale $\Delta^*$ which determines the crossover from ${\cal N}$-channel to 1-channel regime, depends on the precise form of the tunneling amplitudes.
When the off-diagonal elements are small and/or sparse, the onset of the effective one-channel behavior is further delayed, {\it i.e.}, one has to go to lower temperatures to measure $R_q=h/e^2$.


In Fig. \ref{fig:eff_one_channel} we show the variation of the DC conductivity of the multichannel quantum RC circuit, as a function of temperature. The conductivity $G(T)=\frac{4\pi^2e^2}{h}g$, where $g\left(\ln(E_c/T)\right)$ is a dimensionless parameter, has been numerically obtained using the scaling equations in Ref. \onlinecite{Zarand}. We fix the number of channels ${\cal N}={\cal M}=5$ in Fig. \ref{fig:eff_one_channel}, and consider a randomly generated set of tunnel couplings (magenta curve). Using this matrix of tunneling amplitudes, we then restrict ourselves to only the diagonal values (blue curve), setting all the off-diagonal entries to zero. The tunneling amplitudes are globally scaled such that $g(0)=0.5$, corresponding to $T=E_c$. Following this procedure, we then allow off-diagonal diagonal elements representing mixing of adjacent channels (green curve). We then include the elements corresponding to mixing of adjacent and next-to-adjacent channels (red curve), and ultimately additionally include the terms represented cross-mode tunneling of next-to-next-to-adjacent channels (gray curve). From Fig. \ref{fig:eff_one_channel} it can be seen that the system behaves identically to an effective one-channel quantum RC circuit, at low temperatures. Asymmetry in tunneling amplitudes and/or inter-channel mixing are the factors responsible for this emergent one-channel behavior. As apparent from the curves, introducing greater inter-mode couplings, increases the energy scale $\Delta^{\ast}$ (given by the minima of the curves) at which the system crosses over to the one-channel regime. The (black) dashed lines correspond to the unitary limit of the 1-channel Kondo model. The effective Kondo temperature $T_K^{\ast}$\cite{Affleck_2} , is the point at which the curve converges to the unitary limit. For purely diagonal and equal couplings, the system is equivalent to the ${\cal N}$-channel Kondo result. The crossover energy scale $\Delta^{\ast}$ as a function of the number of channels, has been studied \cite{Zarand}, and shown to be suppressed as the number of channels is increased.

The one-channel Kondo fixed point is stable whereas the ${\cal N}$-channel is unstable, and a deviation from the condition $t_{\alpha\beta}=t\delta_{\alpha\beta}$ causes the system to flow away from the ${\cal N}$-channel fixed point. This implies that the quantized resistance
unit $R_q=h/e^2$ may be observable in a multichannel quantum RC circuit, if it is possible to access the energy regime below $\Delta^{\ast}$ experimentally.

%

Deviations from the charge-degeneracy point, {\it i.e.} an effective magnetic field in the pseudospin language, may inhibit the onset of the effective one-channel behavior.  It is a relevant perturbation and the renormalization flow needs to be stopped when the cutoff (temperature) equals the effective magnetic field, a condition determined self-consistently \cite{Zarand}. If this happens before the crossover to the one-channel regime, then the universality of $R_q$ is killed. The variation of $R_q$ for the case of an infinitesimal field, such that the energy scale lies between $T_K$ and $\Delta^{\ast}$, has been studied perturbatively and the one-channel result has been confirmed\cite{Note}. However, the value of $R_q$ precisely at the step, i.e. below $T_K$, remains an open question.

\subsection{Connection with the Anderson-Yuval expansion}

In the previous Subsection we showed that below the energy scale $\Delta^{\ast}$, the multichannel quantum RC circuit near a charge-degeneracy point is equivalent to a one-channel Kondo model. 

Here, we provide a transparent way to better understand the connection of the quantum RC  circuit with the one-channel Kondo model, based on an analogy with the Coulomb gas\cite{Anderson1,Anderson2}.  We assume that a strong Coulomb blockade exists in the quantum dot, and for the sake of simplicity restrict ourselves to the charge-degeneracy point. Deviations from this point, assuming that the Zeeman energy is negligible compared to $E_c$,  can be readily incorporated. However, this makes the mathematical details  significantly more complex without being more illuminating, and will hence be ignored for the sake of clarity. 
Using the Hamiltonian in Eq.\ \eqref{eq:hamiltonian_RC} we can write the partition function as the imaginary-time functional integral
\begin{align}
 Z=\int [{\cal D}\tilde{d}][{\cal D}\tilde{c}]e^{-S},
\end{align}
where the action 
\begin{align}\label{eq:main_action}
&S=\int_{0}^{\beta} d\tau\bigg[\sum_{k\alpha}\bar{\tilde{d}}_{k\alpha}[\partial_{\tau}+\epsilon_{k}]\tilde{d}_{k\alpha}+E_c(\hat{N}-N_{0})^{2}+\notag\\
&\sum_{p\beta}\bar{\tilde{c}}_{p\beta}[\partial_{\tau}+\varepsilon_{p}]\tilde{c}_{p\beta}+t\sum_{k p\alpha\beta}\left(\bar{\tilde{c}}_{p\beta}\tilde{d}_{k\alpha}+\bar{\tilde{d}}_{k\alpha}\tilde{c}_{p\beta}\right)\bigg].
\end{align}
In the above action, we have assumed that we are below the energy scale $\Delta^{\ast}$. In this regime, it is possible to select an appropriate unitary transformation of the degrees of freedom of the system $\{c_{k\alpha},d_{p\beta}\}\rightarrow\{\tilde{c}_{k\alpha},\tilde{d}_{p\beta}\}$, such that the channel-mixing is maximal ($t_{\alpha\beta}=t$) in terms of these effective degrees of freedom.

Introducing the Hubbard-Stratonovich variable $V(\tau)$ via the identity $\int [{\cal D}V]\exp\left[-\int_0^\beta \frac{V^2}{4E_c}d\tau\right]=1$, the quartic term in the $\tilde{d}_{k\alpha}$ operators can be absorbed by the transformation
\begin{align}
 V(\tau)\rightarrow V(\tau)+2iE_{c}(\hat{N}-N_{0}),
\end{align}
such that
 \begin{align}
 Z=\int [{\cal D}\tilde{d}][{\cal D}\tilde{c}][{\cal D} V]e^{-\tilde{S}}. 
\end{align}
Here, the transformed action
\begin{align}
{\tilde S}&= \int_{0}^{\beta} d\tau\bigg[\sum_{k\alpha} \bar{\tilde{d}}_{k\alpha} [\partial_\tau +\epsilon_{k}+iV]\tilde{d}_{k\alpha} +\frac{ V^2}{4 E_c} -i N_0V\notag\\
&+\sum_{p\beta}\bar{\tilde{c}}_{p\beta}[\partial_{\tau}+\varepsilon_{p}]\tilde{c}_{p\beta}+t\sum_{k p\alpha\beta}\left(\bar{\tilde{c}}_{p\alpha}\tilde{d}_{k\beta}+\text{h.c.}\right)\bigg].
\end{align}
The variable $V(\tau)$ can be expanded in terms of its Fourier components $V(\tau)=\frac{1}{\beta}\sum V_{m}e^{i\nu_m \tau}$,
where $\nu_{m}=\frac{2\pi m}{\beta}$ denotes bosonic Matsubara frequencies. The zero mode $V_0$ can be pinned to the values $2n\pi$, where $n\in {\mathbb Z}$, and its fluctuations neglected \cite{Kamenev}. We define the field $\phi_n(\tau)=\int_{0}^{\tau}V_n(\tau')d\tau'$
such that it satisfies the boundary condition $\phi_n(\tau+\beta)=\phi_n(\tau)+2\pi n$. Transforming, 
\begin{eqnarray}
\tilde{d}_{k\alpha}(\tau) &\rightarrow& \tilde{d}_{k\alpha}(\tau)\exp\left[-i\phi_n(\tau)\right] 
\end{eqnarray}
we decouple $N(\tau)$ and $V(\tau)$, while preserving the antiperiodicity $d_{k\alpha}(\tau+\beta)=-d_{k\alpha}(\tau)$. Thus, we obtain
 \begin{align}
 Z=\sum_{n-\infty}^{\infty}\int [{\cal D}\tilde{d}][{\cal D}\tilde{c}][{\cal D} \phi_n]e^{- {\cal S}},
\end{align}
where the transformed action is given by
\begin{align}
{\cal S}& = \int_{0}^{\beta} d\tau\bigg[\sum_{k\alpha} \bar{\tilde{d}}_{k\alpha}[\partial_{\tau}+\epsilon_{k}]\tilde{d}_{k\alpha} +\frac{1}{4 E_c}(\partial_\tau{\phi_n})^2 -i N_0\partial_\tau{\phi_n}\notag\\
&+\sum_{p\beta}\bar{\tilde{c}}_{p\beta}[\partial_{\tau}+\varepsilon_{p}]\tilde{c}_{p\beta}+t\sum_{k p\alpha\beta}\left(\bar{\tilde{c}}_{p\alpha}\tilde{d}_{k\beta}e^{-i\phi_n}+\text{h.c.}\right)\bigg].
\end{align}
Integrating the topological term $N_0\partial_\tau{\phi}$ over (imaginary) time we obtain
\begin{align}
 Z=\sum_{n=-\infty}^{\infty}e^{i2\pi n N_0}\int[{\cal D} \tilde{d}][{\cal D} \tilde{c}][ {\cal D} \phi_n]\exp[-\bf{S}].
\end{align}
We separate the terms in the action as ${\bf S}=S_0^{\text{el}}+S_0^{\phi}+S_{\text{int}}$, where we define
\begin{align}
\label{eq:action_complete}
S_0^{\text{el}} &= \int_{0}^{\beta} d\tau\left[ \sum_{k\alpha} \bar{\tilde{d}}_{k\alpha}[\partial_{\tau}+\epsilon_{k}]\tilde{d}_{k\alpha}+\sum_{p\beta}\bar{\tilde{c}}_{p\beta}[\partial_{\tau}+\varepsilon_{p}]\tilde{c}_{p\beta}\right],
\end{align}
the action for noninteracting electrons in the lead and dot, the kinetic energy of the field $\phi$
\begin{align}
S_0^{\phi}&=\int_{0}^{\beta} d\tau\frac{1}{4 E_c}\left(\frac{\partial\phi_n}{\partial \tau}\right) ^2, 
\end{align}
and finally the interaction term 
\begin{align}
S_{\text{int}}&=\int_{0}^{\beta} d\tau\  t\sum_{k p\alpha\beta}\left(\bar{\tilde{c}}_{p\beta}\tilde{d}_{k\alpha}e^{-i\phi_n}+\text{h.c.}\right).
\end{align}
Expanding the partition function in powers of $S_{\text{int}}$, and subsequently transforming to the charge representation we integrate out the fermionic degrees of freedom. The existence of a strong Coulomb blockade together with the fact that we confine ourselves to the vicinity of the charge-degeneracy point, allows us to project out the other charge states which are energetically distant. This leads to the partition function of the well-known Coulomb gas\cite{Anderson1, Anderson2}
\begin{widetext}
\begin{align}
    Z &=\sum_{l=0}^{\infty} \left(\nu_0\nu_1 t^2\cal N M\right)^l\int_{0}^{\beta}\frac{d\tau_1}{\tau_c}\int_{0}^{\tau_1-\tau_c}\frac{d\tau_2}{\tau_c}\ldots \int_{0}^{\tau_{2l-1}-\tau_c}\frac{d\tau_{2l}}{\tau_c}\exp\left[2\sum_{i<j}(-1)^{i+j}\log\left|\frac{\tau_i-\tau_j}{\tau_c}\right|\right].
    \label{Part}
\end{align}
\end{widetext}
Details of this procedure are included in Appendix \ref{app:B}. Here, the ultraviolet cutoff $1/\tau_c$ is formally identified to be $\Delta^{\ast}$. 

To make a direct connection with the Kondo mapping in the previous Subsection, we could have alternatively rewritten Eq.\ \eqref{eq:main_action} in terms of the totally symmetric mode and ${\cal N}-1$ and ${\cal M}-1$ modes in the dot and lead respectively. The charging energy term is given exclusively in terms of the totally symmetric mode, and the Coulomb repulsion term is what enforces the projection to the two charge states near the charge-degeneracy point. The Coulomb gas expansion follows directly from this fact. This Coulomb gas representation allows us to prove that at energy scales smaller than $E_c$, the effective model for a quantum RC circuit with maximal inter-channel mixing, is the one-channel Kondo model. 

In the complete absence of channel mixing, {\it i.e.} diagonal tunneling matrix elements, rewriting the partition function as a cumulant expansion in $H_\text{T}$ and retaining terms to second-order yields the well-known effective action for the single-electron box in terms of a dissipative particle on a ring\cite{Ambegaokar,Schon,Hofstetter,Larkin,Zakin,Zamo,Lukyanov}. The key ingredient in this mapping is the fact that higher orders in the cumulant expansion are suppressed by factors of at least ${\cal O} (1/{\cal N})$. 

For large-${\cal N}$ this furnishes a natural hierarchy of relevant terms, and in the ${\cal N}\rightarrow\infty$ regime all other terms become completely irrelevant. This is in contrast to the case where channel mixing or anisotropy in the tunneling amplitudes is present, where no such ordering is possible. Defining $ \kappa=\left(\nu_0\nu_1 t^2\cal N M\right)$, all terms in the cumulant expansion in $\kappa$ are of ${\cal O} (1)$ in ${\cal N}$ and ${\cal M}$, once $\kappa$ is held fixed.

\section{Close to perfect transmission}

We now  address the regime opposite to that in Section III, {\it i.e.} a Coulomb-blockaded dot near perfect transmission. In Sub. \ref{perfect_trans} we study the perfect transmission limit, where the system can be exactly solved. We introduce the notation and find the general expression for $R_q$ and $C_0$ for ${\cal N}$ channels. We then focus (Sub. \ref{weak_Kondo}) on the case of 2-channels, where we use an intuitive description of the system in terms of spin and charge degrees of freedom, and introduce weak-backscattering at the dot-lead interface. We extend the results of Ref. 
\onlinecite{Mora_LeHur_2}, which treats the case of perfectly symmetric channels, to include an asymmetry in the backscattering amplitudes. Such an anisotropy in the backscattering amplitudes for a single (orbital) channel of spinful fermions can be introduced by applying an in-plane magnetic field \cite{LeHur_2001}. At high energies we treat perturbatively renormalization effects of the backscattering amplitudes and show that the value of $R_q$ remains unaltered. Lowering the energy cutoff leads to non-universal corrections and we conclude that in the low-energy limit one of the channels is pinched off and the system flows to a single channel Kondo model \cite{LeHur}. Here, $R_q$ equals the unit of quantum resistance, and near the charge-degeneracy point the physics is identical to the one-channel Kondo model obtained in the weak-tunneling regime discussed in Section \ref{weak_tunneling}.

\subsection{Perfectly transmitting channels}\label{perfect_trans}

It is convenient to describe the system using a bosonization approach following Ref. \onlinecite{Flensberg,Matveev2}. The coordinates of this one-dimensional system are chosen such that the lead occupies the semi-infinite line $x\in(-\infty,0)$. The dot extends from $x=0$ to $x=L$. We consider the situation when the level spacing of the dot $\Delta= (\pi v_F)/L$ goes to zero, which implies that $L\rightarrow\infty$. In this limit electrons entering and leaving the dot are uncorrelated. For a concrete analysis, we focus on the case with two conducting channels. We bosonize the total Hamiltonian $H=H_{\text k}+H_{\text c}+H_{\text{bs}}$ \cite{Giamarchi} (for an excellent survey on the subject and details of the notation used, see Ref. \onlinecite{Aleiner}). Here, we assume that the channels are perfectly tansmitting. The role of backscattering will be investigated in the next subsection. 

The kinetic term takes the form \cite{Matveev2,LeHur}:
\begin{equation}
H_{\text k} = \frac{v_F}{2\pi} \int_{-\infty}^{+\infty} dx \sum_{i=1,2} [(\partial_x \phi_i(x))^2 +(\partial_x\theta_{i}(x))^2].
\end{equation}
Here, $v_F$ is the Fermi velocity which is obtained by linearizing the energy spectrum around the Fermi points. The ultraviolet cutoff $1/a$ denotes the extent around $k_F$ upto which the spectrum can be linearized. We note that the bosonic field operators $\phi_i(x)$ and $\theta_i(x)$ obey the usual commutation relations:
\begin{equation}
[\phi_i(x),\theta_j(y)] = \frac{i\pi }{2}\hbox{sgn}(x-y)\delta_{ij}.
\end{equation}
In particular, the total electron charge on the cavity becomes $\hat{Q}=(1/\pi)(\phi_1(0) +\phi_2(0))$ such that the charging Hamiltonian takes the form \cite{Matveev2,LeHur}:
\begin{equation}
H_{\text c} = \frac{E_c}{\pi^2}\left(\sum_{i=1,2} \phi_i(0) - \pi N_0\right)^2,
\end{equation} 
and $N_0$, which is related to the gate voltage, has been introduced earlier in Sec. II. For electrons in the cavity, we choose $\phi_i(\infty)=0$ which fixes the charge on the dot to be zero when $N_0=0$. Here, the two conducting modes can refer to the two spin polarizations of the electron or two electron channels. In either case it is then natural to introduce the charge and (pseudo-)spin modes $\phi_{c,s}=(\phi_1(0)\pm \phi_2(0))/\sqrt{2}$. 

At perfect transmission, we can omit the spin part of the Hamiltonian and integrate out the charge fields except the one at $x=0$, which results in the following action:
\begin{equation}
S_0^{c} =  \sum_{\omega_n}\left|\hat{\phi}_c(0,\omega_n)\right|^2\left(|\omega_n| +\frac{2 E_c}{\pi}\right).
\label{eq:action_origin}
\end{equation}
Here, we have defined the shifted charge field $\hat{\phi_c}(0,\omega_n)=\phi_c(0,\omega_n)-\pi N_0$, in order to make the action independent of $N_0$. From the Green's function of the charge mode $\phi_c(0,\omega_n)$, one can easily evaluate the function $K(t)$ in Eq. (\ref{K}) and obtain
\begin{equation}
\frac{\hat{Q}}{V_g} = \frac{C_g}{1-\frac{i\omega \pi}{2 E_c}}.
\label{eq:K0_reference}
\end{equation}
Above, we have chosen units in which $\hbar=1$. Comparing the low-frequency expansion of Eq. \eqref{eq:K0_reference} with Eq. (\ref{charge}) this results in:
\begin{equation}
R_q = \frac{h}{2e^2}.
\end{equation}
At a general level, for a large cavity and a number ${\cal N}$ of perfectly conducting channels one finds $R_q=h/({\cal N} e^2)$.
In Section III we showed that the presence of channel mixing and/or asymmetry in the tunneling amplitudes cause $R_q$ to be universal and equals the unit of resistance $h/e^2$. In the remainder of this section we study the effect of backscattering on $R_q$; in particular, we investigate the role of asymmetry of the backscattering amplitudes on the relaxation resistance. 

\subsection{Weak backscattering and 2-channel Kondo physics}\label{weak_Kondo}

A deviation from the perfect transmission condition is captured by the back-scattering Hamiltonian \cite{LeHur}
\begin{align}
H_{\text{bs}}=\frac{1}{\pi a}\sum_{i=1,2}\left|\lambda_i\right|\cos\left(2 \phi_i(0)\right).
\end{align}
In terms of the charge and spin fields this can be recast as
\begin{align}
H_{\text{bs}}=\frac{1}{\pi a}\sum_i\left|\lambda_i\right|\cos\left(\sqrt{2}\left(\phi_c+\eta_i\phi_s\right)\right),
\label{eq:intermediate}
\end{align}
where we have introduced $\eta_1=1$ and $\eta_2=-1$. Defining $\lambda= \left|\lambda_1\right|+\left|\lambda_2\right|$ and $\delta_\lambda= \left|\lambda_2\right|-\left|\lambda_1\right|$  we can rewrite Eq. \eqref{eq:intermediate} as
\begin{align}
H_{\text{bs}}=&\frac{1}{\pi a}\bigg[\lambda\cos\left(\sqrt{2}\hat{\phi}_c-\pi N_0\right)\cos\left(\sqrt{2}\phi_s\right)\notag\\
&\qquad+\delta_\lambda\sin\left(\sqrt{2}\hat{\phi_c}-\pi N_0\right)\sin\left(\sqrt{2}\phi_s\right)\bigg].
\end{align}
In the Coulomb blockade regime, the shifted field $\hat{\phi}_c$ is pinned to to zero. To be mathematically rigorous we adopt a coherent state functional integral approach in the ensuing treatment. 
The reader is to assume that the various transformations are implemented in the action and not at the level of operators. 

We define the energy scales $\Gamma_{+}=\frac{2}{v_F}\left(\frac{\lambda \cos(\pi N_0)}{\sqrt{2\pi a}}\right)^2$ and $\Gamma_{-}=\frac{2}{v_F}\left(\frac{\delta_{\lambda} \sin(\pi N_0)}{\sqrt{2\pi a}}\right)^2$. A physical interpretation of these energies in terms of lifetimes of Majorana fermions is discussed later in this subsection.  To study the low-energy physics, following Ref. \onlinecite{Mora_LeHur_2}, we introduce an intermediate energy scale $\Lambda$, such that $\max\{\Gamma_+,\Gamma_-\}\ll\Lambda\ll E_c$. We then integrate out the high-energy contribution $\Lambda<\omega< E_c$ to the charge mode, such that the resultant action still effectively describe the interaction between charge and spin modes at the energy scales of $\Gamma_{\pm}$.  This amounts to replacing the factor $\cos\left(\sqrt{2}\hat{\phi}_c-\pi N_0\right)$ with 
\begin{align}
\cos\left(\sqrt{2}\hat{\phi}^{l}_c-\pi N_0\right)e^{- \langle\hat{\phi}_c(\tau)\hat{\phi}_c(\tau)\rangle_{\epsilon>\Lambda}},
\end{align}
where $\hat{\phi}^{l}_c$ contains the low-energy fluctuations of the charge field.
In this regime, we can expand the backscattering terms in powers of $\hat{\phi}^{l}_c$, and approximate
\begin{align}
\cos\left(\sqrt{2}\hat{\phi}^{l}_c-\pi N_0\right)&\approx \cos(\pi N_0)+\sqrt{2}\sin(\pi N_0)\hat{\phi}^{l}_c\notag\\
-\cos(\pi N_0)&(\hat{\phi}^{l}_c)^2:= \cos(\pi N_0)+\sqrt{2\pi a}\hat{C}_+\\
\sin\left(\sqrt{2}\hat{\phi}^{l}_c-\pi N_0\right)&\approx -\sin(\pi N_0)+\sqrt{2}\cos(\pi N_0)\hat{\phi}^{l}_c\notag\\
+\sin(\pi N_0)&(\hat{\phi}^{l}_c)^2:= -\sin(\pi N_0)+\sqrt{2\pi a}\hat{C}_-.
\end{align} 
We define the dimensionless backscattering amplitudes $r_{1,2}$ via the relation
\begin{align}
\lambda_{1,2}=r_{1,2}\sqrt{\frac{ 2 a v_F E_C\gamma}{\pi}}.
\end{align}
 In other words we set $\langle\hat{\phi}_c(\tau)\hat{\phi}_c(\tau)\rangle_{\epsilon>\Lambda}\simeq (1/2)\ln(\omega_D\pi/2E_c)$, where $\omega_D=v_F/\gamma a$, $\gamma$ being the Euler-Mascheroni constant. 

In the ensuing analysis we implicitly assume $\hat{\phi}^{l}_c$ to be small. Since the charging energy $E_c$ is assumed to be much larger than the backscattering amplitudes, the dynamics of the spin field are completely captured by the pinned charge field $\hat{\phi}^{l}_c$. Next, we compute corrections to $R_q$ due to charge fluctuations and study the role of an asymmetry in the backscattering amplitides. 

The spin part of the Hamiltonian can be treated exactly by refermionizing the field $\phi_s$ using the procedure outlined in Ref. \onlinecite{LeHur}.  The kinetic part of the Hamiltonian can be written as the sum of its charge and spin parts, {i.e.} $H_{\text{k}}=H^{c}_{\text{k}}+H^{s}_{\text{k}}$. Rewriting $H^{c}_{\text{k}}$ as a functional integral and integrating out the charge field away from $x=0$ we obtain the action in Eq. \eqref{eq:action_origin}. In terms of the new fermionic fields $\psi_s(x)$ and the ``slave'' fermion $d$, the kinetic part of the spin Hamiltonian becomes
\begin{align}
H^{s}_{\text{k}}=-iv_F\int_{-\infty}^{\infty} dx \psi_s^{\dag}(x)\partial_x \psi_s(x),
\label{eq:spin_refermionized}
\end{align}
and the backscattering Hamiltonian is given by
\begin{align}
H_{\text{bs}}=&\bigg[(X_{+}+\lambda\hat{C}_+)\left(\psi^{\dag}_s(0)+\psi_s(0)\right)(d-d^{\dag})\notag\\
&+(X_{-}+\delta_{\lambda}\hat{C}_{-})\left(\psi^{\dag}_s(0)-\psi_s(0)\right)(d+d^{\dag})\bigg],
\label{eq:backscattering_refermionized}
\end{align}
reminiscent of the two-channel Kondo model \cite{Emery-Kivelson,Cox} in the presence of a channel asymmetry \cite{Fabrizio,LeHur}. The introduction of two Majorana fermions follows from Ref. \onlinecite{LeHur}, and Ref. \onlinecite{Fabrizio} uses this representation to make an analogy with the 2-channel Kondo model in the presence of a channel asymmetry (at the Emery-Kivelson line).This Majorana model also finds applications in the context of dissipative mesoscopic structures \cite{Finkelstein}. 

Here, we have defined $X_+=\lambda\cos(\pi N_0)/\sqrt{2\pi a}$ and $X_{-}=-\delta_\lambda\sin(\pi N_0)/\sqrt{2\pi a}$, to isolate the terms which involve the shifted charge field $\hat{\phi}^{l}_c$. These terms capture the dependence of the spin propagator on the pinned charged and is included in the unperturbed Hamiltonian. Next, the action of the remainder of $H_{\text{bs}}$ on the charge response function is treated perturbatively. For this purpose it is convenient to recast this Hamiltonian in terms of Majorana fermions on the lines of Ref. \onlinecite{LeHur}. Details of this computation are provided in Appendix \ref{app:C}. 

 We define the functions $\kappa_+(N_0):=\cos(\pi N_0)$ and $\kappa_-(N_0):=\sin(\pi N_0)$. The energy scale of the Fermi velocity, which is lost in the bosonization process where an infinite bandwidth is assumed, is reintroduced by fixing the cutoff \cite{Flensberg}. The widths $\Gamma_{+}$ and $\Gamma_{-}$ of the impurity Majoranas, given by $\Gamma_{\zeta} = 2X_{\zeta}^2/v_F$, determine the two intrinsic energy scales in the problem, corresponding to the total backscattering strenth and the channel asymmetry respectively; they can be explicitly obtained through the Renormalization Group (RG) approach  \cite{LeHur}. We define the infrared cutoff $\Gamma=\max\{\Gamma_+,\Gamma_-\}$. It corresponds to the energy scale which is inversely proportional to the lifetime of the (shorter lived) Majorana fermion. As previously stated, the system can be mapped onto the 2-channel Kondo model at the Emery-Kivelson line \cite{Emery-Kivelson} and in terms of this they represent the Kondo coupling constants $J_x$ and $J_y$ respectively. 
 
 The expression for the charge reponse function to leading order in the backscattering can be written as 
\begin{align}
K(\omega)=K_0(\omega)+K_1(\omega)+K_2(\omega)+O(\lambda^4),
\end{align}
where 
\begin{widetext}
\begin{align}
K_1(\omega)&=\frac{\gamma}{2\pi E_c}\frac{1}{\left(1-i\frac{\pi\omega}{2E_c}\right)^2} \left[\sum_{\zeta= \pm} \kappa_{\zeta}(N_0)^2\left(r_1 + \zeta r_2\right)^2\right]\log\left(\frac{E_c}{\Gamma}\right)\notag\\
K_2(\omega)&=-\frac{\gamma}{2\pi E_c}\frac{1}{\left(1-i\frac{\pi\omega}{2E_c}\right)^2}\left[\sum_{\zeta = \pm} \kappa_{-\zeta}(N_0)^2 \left(r_1 + \zeta r_2\right)^2\right]\left(\log\left(\frac{E_c}{\Gamma}\right)-\left(1+i\frac{2\Gamma}{\omega}\right)\log\left(1-i\frac{\omega}{\Gamma}\right)\right).
\end{align}
\end{widetext}
When the backscattering amplitudes are symmetric the contributions corresponding to $\zeta=-$ disappear and this case has been looked at in detail in Ref. \onlinecite{Mora_LeHur_2}. The level asymmetry, which is captured by the energy scale $\Gamma_{-}$ plays a crucial role in the low energy behavior of the system which we discuss below. 

We now compare
\begin{align}
K(\omega)=i \frac{2}{\pi^2}\int_{-\infty}^{\infty}dt e^{i\omega t}\left\langle \left\{\hat{\phi}_c(\tau),\hat{\phi}_c(0)\right\}\right\rangle
\end{align}
with the expansion $K(\omega)=C_0(1+i\omega C_0 R_q)+O\left(\omega^2\right)$. To proceed systematically we formally expand the functions $K_0(\omega)$, $K_1(\omega)$ and $K_2(\omega)$ to linear order in the frequencies
\begin{align}
    K_0(\omega) &= A_0 + i \omega B_0 \notag \\
    K_1(\omega) &= \sum_{\zeta=\pm}r_{\zeta}^2\left[ A^{\zeta}_1 + i \omega B^{\zeta}_1\right] \notag \\
    K_2(\omega) &=  \sum_{\zeta=\pm}r_{\zeta}^2\left[A^{\zeta}_2 + i \omega B^{\zeta}_2\right].
\end{align}
From this, we find the mesoscopic capacitance 
\begin{align}\label{eq:Rpert}
    C_0 &= A_0 + \sum_{\zeta=\pm}r_{\zeta}^2\left[ A^{\zeta}_1 + A^{\zeta}_2\right] \notag \\
    R_q &=  \frac{(B_0 + \sum_{\zeta=\pm}r_{\zeta}^2\left[ B^{\zeta}_1 +  B^{\zeta}_2\right]}{(A_0 + \sum_{\zeta=\pm}r_{\zeta}^2\left[ A^{\zeta}_1 + A^{\zeta}_2\right])^2}
\end{align}
where we have retained terms to leading order in the backscattering amplitudes. Here, we have defined the parameters $r_\pm=r_1\pm r_2$. 

One then finds to $O(r_{\zeta}^2)$ 
\begin{align}
\label{eq:Rq_bs}
C_0&=\frac{1}{2E_c}-\frac{\gamma}{\pi E_c}\sum_{\zeta=\pm} r_{\zeta}^2\kappa_{-\zeta}(N_0)^2\notag\\
&\qquad+\frac{\gamma}{2\pi E_c}\sum_{\zeta=\pm} \zeta r_{\zeta}^2\cos(2\pi N_0)\log\left(\frac{E_c}{\Gamma}\right)\notag\\
R_q&=\frac{h}{2e^2}.
\end{align}
In the vicinity of $N_0=1/2$ the logarithm term dominates and in particular setting $r_1=r_2$ one recovers the correction to the mesoscopic capacitance $\delta C= 2\gamma/(\pi E_c)r^2\cos(2\pi N_0)\log\left(1/(r^2\cos^2(\pi N_0))\right)$ obtained in Ref. \onlinecite{Matveev}. Specifically, the logarithmic contribution of the capacitance $C_0$ agrees with Ref. \onlinecite{LeHur} in the presence of a channel asymmetry. Note, $R_q$ is not affected by backscattering and remains $h/2e^2$ to the second order in the backscattering amplitudes. Non-universal corrections to $R_q$ nevertheless appear at fourth or higher order in backscattering amplitudes. 

At low frequencies and temperatures the backscattering amplitudes are substantially renormalized, thus entering the regime of strong backscattering, and this perturbative scheme to compute $R_q$ is inadequate.
For low energies, the channel asymmetry for energy scales below $\Gamma_{-}$ grows till $r_{\text{max}}=\max\{r_1,r_2\}\rightarrow 1$, indicating that one of the channels is pinched off and there is only one channel effectively contributing to transport \cite{LeHur}. In this regime, near a charge degeneracy point, the system is described by a 1-channel Kondo model and the value $h/e^2$ of $R_q$ should reemerge in agreement with the weak-tunneling limit.

\section{Conclusion}
To summarize, we have investigated dissipation effects in a strongly correlated multichannel quantum RC circuit through the charge dynamics at low frequency and the concept of charge relaxation resistance. We have corroborated the violation of the Kirchhoff's law as a result of quantum coherence effects. Our main message is that in the vicinity of a charge degeneracy point and in the case of a large cavity, the system can flow to a one-channel Kondo fixed point at low energy as a result of channel mixing or channel asymmetry, leading to a resistance unit $h/e^2$ close to absolute zero reminiscent of the one-channel RC circuit \cite{ChristKaryn}. 

It is relevant to note that such a resistance unit has also appeared in a distinct purely non-equilibrium context of Landau-Zener sweep on the gate voltage \cite{Etzioni1,Etzioni2}. In the latter situation, the effect of channel mixing or channel asymmetry was not considered which suggests a different origin for the emergence of the resistance quantum. 

It is also important to underline that for the multichannel quantum RC circuit, prominent deviations from the resistance quantum are expected either in regions where the weak backscattering or weak tunneling limit describing the coupling between lead and cavity is applicable. We believe that our results shed new light on the understanding of the dynamics of quantum RC circuits in close connection with multichannel quantum impurity models \cite{NozieresBlandin,CoqblinSchrieffer}, and could be in principle detected with current technology \cite{Berman,Lehnert}, especially close to perfect transmission \cite{Goldhaber}.

%
%

We thank B. Horovitz and P. Le Doussal for fruitful discussions.  This work is supported by Department of Energy under the grant DE-FG02-08ER46541 (P.D and K.L.H).

\appendix
\section{\label{app:A}Perturbation Theory in weak tunneling regime}

In this Appendix, we present the derivation related to the main Eq. (\ref{Rqb}) in Sec. II. We start from the function $K(t)$ defined in Eq. (\ref{K}) and first we focus
on the second order contribution in the tunneling amplitude, which will generate the mesoscopic capacitance $C_0$ \cite{Matveev}. We generalize the computation of Ref. \onlinecite{ChristKaryn} which has been performed in the context of the one-channel quantum RC circuit.

\subsection{2nd order}

Assuming $t>0$ the response function can be simplified to give
\begin{align}
 K^{(2)}(t)&=-i\int_{t}^{\infty} dt_{1}\int_{-\infty}^{0}dt_{2}\notag\\
 &\times\langle\phi_{GS}|H_\text{T}(t_1) \hat{N}(t)\hat{N}(0)H_\text{T} (t_2)|\phi_{GS}\rangle.
\label{eq:inter}
\end{align}

The contributions can be conveniently indexed by the path adopted by the dot electrons. There are two possible paths which give non-zero contributions. $0\rightarrow 1\rightarrow 0$ represents a lead electron hopping onto the dot and back, while  $0\rightarrow -1\rightarrow 0$ represents a dot electron hopping onto the lead and back.

\noindent For $t>0$ the contribution of the $0\rightarrow 1\rightarrow 0$ path is evaluated to be 
\begin{align}
&\sum_{kp\alpha\beta}t_{\alpha\beta}^2\int_{t}^{\infty} dt_{1}\int_{-\infty}^{0}dt_{2}e^{-i\left[\epsilon_{k}-\varepsilon_{p}+E_c(1-2N_0)\right]\left(t_1-t_2\right)}\notag\\
&\qquad\qquad\times\theta(-\varepsilon_p)\theta(\epsilon_k).
\end{align}
Similarly, the  $0\rightarrow -1\rightarrow 0$ path contributes
\begin{align}
&\sum_{kp\alpha\beta}t_{\alpha\beta}^2\int_{t}^{\infty} dt_{1}\int_{-\infty}^{0}dt_{2}e^{-i\left[\varepsilon_{p}-\epsilon_{k}+E_c(1+2N_0)\right]\left(t_1-t_2\right)}\notag\\
&\qquad\qquad\times\theta(\varepsilon_p)\theta(-\epsilon_k).
\end{align}
Collecting these contributions together, we obtain for ($t>0$)
\begin{align}
K^{(2)}(t)&=i\sum_{kp\alpha\beta}t_{\alpha\beta}^2\bigg[\frac{\theta(\epsilon_{k})\theta(-\varepsilon_{p})}{\Delta_1^2}e^{-i\Delta_{1}(k,p)t}\notag\\
&\qquad\qquad\qquad+\frac{\theta(-\epsilon_{k})\theta(\varepsilon_{p})}{\Delta_{-1}^2}e^{-i\Delta_{-1}(k,p)t}\bigg]
\label{eq:t_g_0}.
\end{align}
Here, we have defined
\begin{align}
 \Delta_{\pm 1}(k,p)=E_{c}(1\mp2N_{0})\pm(\epsilon_{k}-\varepsilon_{p}). 
\end{align}
The $t<0$ contribution can be deduced from Eq. \eqref{eq:t_g_0} by the replacement $t\rightarrow -t$. Fourier transforming  we obtain
\begin{align}
K^{(2)}(\omega)&=\int_{-\infty}^{\infty}dt e^{i\omega t}e^{-0^{+}|t|}K^{(2)}(t)\notag\\
&=\sum_{kp\alpha\beta}t_{\alpha\beta}^2\sum_{\sigma=\pm 1}\frac{\theta(-\varepsilon_{p})\theta(\epsilon_{k})}{\left[\Delta_{\sigma}(k,p)\right]^{2}}\bigg(\frac{1}{\Delta_{\sigma}(k,p)
+\omega+i\eta}\notag\\
&\qquad\qquad+\frac{1}{\Delta_{\sigma}(k,p)-\omega+i\eta}\bigg).
\end{align}
Assuming a metallic dot, we can let $\sum_k\rightarrow\nu_1 \int_{-\infty}^{\infty}d\epsilon_k$ (for the lead we have  $\sum_p\rightarrow\nu_0 \int_{-\infty}^{\infty}d\varepsilon_p$) and the mesoscopic capacitance is determined to be
\begin{align} 
C_{0}&=e^{2} K^{(2)}(0)=\frac{\nu_{0}\nu_{1}C_{g}}{1/4-N_{0}^{2}}\sum_{\alpha\beta}t_{\alpha\beta}^{2}.
\label{eq:second_C_0}
\end{align}
This corresponds to Eq. (\ref{C0}) in the main text. This shows that in the weak-tunneling regime, the mesoscopic capacitance of a single-electron box differs
very strongly from the geometrical capacitance as a result of the Coulomb blockade phenomenon.

\subsection{4th order}

The various contributions which arise at 4th order are classified by indexing the dynamics of the charge occupation on the dot and the number of virtual particle-hole excitations. Only the paths $0\rightarrow 1\rightarrow 0\rightarrow 1\rightarrow0$, $0\rightarrow 1\rightarrow 0\rightarrow -1\rightarrow0$, $0\rightarrow -1\rightarrow 0\rightarrow 1\rightarrow0$ and $0\rightarrow -1\rightarrow 0\rightarrow -1\rightarrow0$ contribute at low energies ({\it i.e.} linear in $\omega$). Furthermore, based on phase space arguments, only paths within this subset that have particle-hole virtual excitations contribute. 
We explicitly compute the relevant contribution by the path $0\rightarrow 1\rightarrow 0\rightarrow 1\rightarrow0$. Its contribution to  $K^{(4)}$ for $t>0$ follows from the expression
\begin{align}
&\int_{t}^{\infty}dt_{1}\int_{0}^{t}dt_{2}\int_{0}^{t_{2}}dt_{3}\int_{-\infty}^{0}dt_{4}\langle\phi_{GS}|H_\text{T}(t_1) \hat{N}(t)H_\text{T} (t_2)\notag\\
&\quad\times H_\text{T} (t_3)\hat{N}(0)H_\text{T} (t_4)|\phi_{GS}\rangle.
\end{align}
Isolating the product of operators contributing to this path we obtain
\begin{align}
&\sum_{\genfrac{}{}{0pt}
{}{k_1 p_1 \alpha_1\beta_1}{\ldots \beta_4}}t_{\alpha_1\beta_1}t_{\alpha_2\beta_2}t_{\alpha_3\beta_3}t_{\alpha_4\beta_4}\int_{t}^{\infty}dt_{1}\int_{0}^{t}dt_{2}\int_{0}^{t_{2}}dt_{3}\int_{-\infty}^{0}dt_{4}\notag\\
&\times e^{-i\left(\Delta_{1}(k_{1},p_{1})t_{1}-\Delta_{1}(k_{2},p_{2})t_{2}+\Delta_{1}(k_{3},p_{3})t_{3}-\Delta_{1}(k_{4},p_{4})t_{4}\right)}\notag\\
&\times\langle\phi_{GS}|c^{\dag}_{p_{1}\beta_{1}}d_{k_{1}\alpha_{1}}d^{\dag}_{k_{2}\alpha_{2}}c_{p_{2}\beta_{2}} c^{\dag}_{p_{3}\beta_{3}}d_{k_{3}\alpha_{3}}d^{\dag}_{k_{4}\alpha_{4}}c_{p_{4}\beta_{4}}|\phi_{GS}\rangle.
\end{align}
The pairings corresponding to single particle-hole excitations in the intermediate state are given by
\begin{align}
 \contraction[2ex]{\langle\phi_{GS}|}{c}{^{\dag}_{p_{1}\beta_{1}}d_{k_{1}\alpha_{1}}d^{\dag}_{k_{2}\alpha_{2}}c_{p_{2}\beta_{2}} c^{\dag}_{p_{3}\beta_{3}}d_{k_{3}\alpha_{3}}d^{\dag}_{k_{4}\alpha_{4}}}{c}
 \contraction{\langle\phi_{GS}|c^{\dag}_{p_{1}\beta_{1}}}{d}{_{k_{1}\alpha_{1}}}{d}
 \contraction{\langle\phi_{GS}|c^{\dag}_{p_{1}\beta_{1}}d_{k_{1}\alpha_{1}}d^{\dag}_{k_{2}\alpha_{2}}}{c}{_{p_{2}\beta_{2}}}{c}
 \contraction{\langle\phi_{GS}|c^{\dag}_{p_{1}\beta_{1}}d_{k_{1}\alpha_{1}}d^{\dag}_{k_{2}\alpha_{2}}c_{p_{2}\beta_{2}} c^{\dag}_{p_{3}\beta_{3}}}{d}{_{k_{3}\alpha_{3}}}{d}
\langle\phi_{GS}|c^{\dag}_{p_{1}\beta_{1}}d_{k_{1}\alpha_{1}}d^{\dag}_{k_{2}\alpha_{2}}c_{p_{2}\beta_{2}} c^{\dag}_{p_{3}\beta_{3}}d_{k_{3}\alpha_{3}}d^{\dag}_{k_{4}\alpha_{4}}c_{p_{4}\beta_{4}}|\phi_{GS}\rangle
\label{eq:eh_lead}
\end{align}
and
\begin{align}
 \contraction{\langle\phi_{GS}|}{c}{^{\dag}_{p_{1}\beta_{1}}d_{k_{1}\alpha_{1}}d^{\dag}_{k_{2}\alpha_{2}}}{c} 
\contraction[2ex]{\langle\phi_{GS}|c^{\dag}_{p_{1}\beta_{1}}}{d}{_{k_{1}\alpha_{1}}d^{\dag}_{k_{2}\alpha_{2}}c_{p_{2}\beta_{2}} c^{\dag}_{p_{3}\beta_{3}}d_{k_{3}\alpha_{3}}}{d}
\contraction[3ex]{\langle\phi_{GS}|c^{\dag}_{p_{1}\beta_{1}}d_{k_{1}\alpha_{1}}}{d}{^{\dag}_{k_{2}\alpha_{2}}c_{p_{2}\beta_{2}} c^{\dag}_{p_{3}\beta_{3}}}{d}
\contraction{\langle\phi_{GS}|c^{\dag}_{p_{1}\beta_{1}}d_{k_{1}\alpha_{1}}d^{\dag}_{k_{2}\alpha_{2}}c_{p_{2}\beta_{2}}}{c}{^{\dag}_{p_{3}\beta_{3}}d_{k_{3}\alpha_{3}}d^{\dag}_{k_{4}}\alpha_{4}}{c}
\langle\phi_{GS}|c^{\dag}_{p_{1}\beta_{1}}d_{k_{1}\alpha_{1}}d^{\dag}_{k_{2}\alpha_{2}}c_{p_{2}\beta_{2}} c^{\dag}_{p_{3}\beta_{3}}d_{k_{3}\alpha_{3}}d^{\dag}_{k_{4}\alpha_{4}}c_{p_{4}\beta_{4}}|\phi_{GS}\rangle.
\label{eq:eh_dot}
\end{align}
The pairing \eqref{eq:eh_lead} corresponds to an electron-hole excitation in the reservoir, whereas \eqref{eq:eh_dot} corresponds to an electron-hole excitation in the dot.
Integrating over the time variables and extracting the  imaginary contribution of the above expression leads to a contribution linear in $\omega$ (hence gapless)
\begin{align}
&\sum_{\genfrac{}{}{0pt}
{}{k_1 p_1 k_2 p_2}{\alpha_1\beta_1\alpha_2 \beta_2}}t_{\alpha_1\beta_1}t_{\alpha_2\beta_1}t_{\alpha_2\beta_2} t_{\alpha_1\beta_2}\theta\left(\epsilon_{k_{1}}\right)\theta\left(\epsilon_{k_{2}}\right)\theta\left(-\varepsilon_{p_{1}}\right)\theta\left(\varepsilon_{p_{2}}\right)\notag\\
&\times\frac{e^{i\left[\Delta_{1}(k_{1},p_{2})-\Delta_{1}(k_{1},p_{1})\right]t}}{\Delta_{1}(k_{1},p_{1})\Delta_{1}(k_{2},p_{1})\Delta_{1}(k_{2},p_{2})\Delta_{1}(k_{1},p_{2})}
\end{align}
The contribution of this term to the response function $K(\omega)$ to linear order in $\omega$ is evaluated to be 
\begin{align}
\Im m  K^{(4)}_{01010}&(\omega)=i\pi\nu_0^2\omega \sum_{\genfrac{}{}{0pt}
{}{\alpha_1\beta_1}{\alpha_2 \beta_2}}t_{\alpha_1\beta_1}t_{\alpha_2\beta_1}t_{\alpha_2\beta_2} t_{\alpha_1\beta_2}\notag\\
\times&\left[\sum_{k}\frac{\theta(\epsilon_{k})}{[E_c(1-2N_0)+\epsilon_k]^2}\right]^2+{\cal O}(\omega^2).
\end{align}

Note that the imaginary contribution for $\omega<0$ arises from the $t<0$ expression. Collecting the contribution from the paths $0\rightarrow1\rightarrow0\rightarrow-1\rightarrow0$, $0\rightarrow-1\rightarrow0\rightarrow1\rightarrow0$ and $0\rightarrow-1\rightarrow0\rightarrow-1\rightarrow0$ in a similar fashion, and
 noting that for electron-hole excitations in the lead given by the grouping in Eq.\ \eqref{eq:eh_lead} we obtain an identical contribution
\begin{align}
 \Im m K^{(4)}&(\omega)=2\pi\nu_0^2\nu_1^2\omega \sum_{\genfrac{}{}{0pt}
{}{\alpha_1\beta_1}{\alpha_2\beta_2}}t_{\alpha_1\beta_1}t_{\alpha_2\beta_1}t_{\alpha_2\beta_2} t_{\alpha_1\beta_2}\notag\\
\times&\left[\frac{1}{E_c(1-2N_0)}+\frac{1}{E_c(1+2N_0)}\right]+{\cal O}(\omega^2).
\label{eq:4th_order_rs}
\end{align}
In the last step we assume a metallic dot and let $\sum_k\rightarrow\nu_1 \int_{-\infty}^{\infty}d\epsilon_k$ (for the lead we have  $\sum_p\rightarrow\nu_0 \int_{-\infty}^{\infty}d\varepsilon_p$). Comparing Eq.\ \eqref{eq:4th_order_rs} with the general relation $ \frac{Q(\omega)}{V_{g}(\omega)}=C_{0}(1+i\omega C_{0}R_{q})+{\cal O}(\omega^{2})$ and using Eq.\ \eqref{eq:second_C_0} we identify the AC charge relaxation resistance to be
\begin{align}
 R_{q}=\frac{2\pi\hbar}{e^{2}}\left[\frac{ \sum_{\genfrac{}{}{0pt}
{}{\alpha_1\beta_1}{\alpha_2\beta_2}}t_{\alpha_1\beta_1}t_{\alpha_2\beta_1}t_{\alpha_2\beta_2} t_{\alpha_1\beta_2}}{\left(\sum_{\alpha\beta}t_{\alpha\beta}^{2}\right)^{2}}\right].
\end{align}
This corresponds to the formula (\ref{Rqb}) in the main text. Note that here we have explicitly restored the Planck constant $\hbar$ which was set to one in the
derivation.

\section{\label{app:B}Coulomb gas formulation}

Here, we show that for energy scales below $\Delta^*$, the partition function close to the charge degeneracy points can be rewritten exactly as the partition function of the one-channel Kondo model, resorting to the Coulomb gas analogy \cite{Anderson1,Anderson2}. By assuming that only one mode, {\it e.g.},  the totally symmetric mode penetrates into the cavity for energy scales below $\Delta^*$, Eq. (\ref{Kondo}) formally reproduces a one-channel Kondo model. Here, the goal of this Appendix is to provide an alternative derivation, following the same terminology as the one used in Refs.~\onlinecite{Ambegaokar,Schon,Hofstetter,Larkin,Zakin,Zamo,Lukyanov}, but keeping the channel-mixing terms. 

First, it is relevant to observe that the partition function in Eq.\ \eqref{eq:action_complete} can be expanded in powers of $t$ as follows
\begin{align}
 Z&=\sum_{n=-\infty}^{\infty}e^{i2\pi n N_0}\int[{\cal D} d][{\cal D} c][ {\cal D} \phi_n]\exp[-S_0^{\text{el}}-S_0^{\phi}]\notag\\
 &\qquad\qquad\times\sum_{m=0}^{\infty}\frac{(-1)^m}{m!}\left(S_{\text{int}}\right)^m\notag\\
 &=\sum_{n=-\infty}^{\infty}\sum_{m=0}^{\infty} \frac{(-1)^m}{m!} e^{i2\pi n N_0}\int[ {\cal D} \phi]e^{-S_0^{\phi}}\left\langle\left( S_{\text{int}}\right)^m\right\rangle^{\text{el}}_0.
\end{align}
The notation $\left\langle\ldots\right\rangle^{\text{el}}_0$ implies that the the expectation value is taken with respect to the (free) actions of the electrons in the lead and dot. 

Let us simplify the term $\left\langle\left( S_{\text{int}}\right)^m\right\rangle^{\text{el}}_0$.
\begin{widetext}
\begin{align}
 \left\langle\left( S_{\text{int}}\right)^m\right\rangle^{\text{el}}_0=t^m\int_{0}^{\beta}d\tau_1\ldots\int_{0}^{\beta}d\tau_m\sum_{k_1 p_1 \alpha_1\beta_1}\ldots\sum_{k_m p_m \alpha_m\beta_m}\left\langle \left(\bar{c}_{p_1\beta_1}d_{k_1\alpha_1}e^{-i\phi_1}+\text{h.c.}\right)\ldots \left(\bar{c}_{p_m\beta_m}d_{k_m\alpha_m}e^{-i\phi_m}+\text{h.c.}\right)\right\rangle.
\end{align}
\end{widetext}
We alter the notation, such that the subscript $\phi_k=\phi(\tau_k)$ is now taken to indicate the time label of the $\phi$ operator. It is implicitly assumed that the 
$\phi$ fields obey twisted boundary conditions. This expression has contributions only when $m=2l$ is even, and furthermore contains an equal number of $c$ and $\bar{c}$ operators (and as a corollary $d$ and $\bar{d}$ operators). The most general term looks like 
\begin{widetext}
\begin{align}
 {\cal A}\ =&t^{2l}\int_{0}^{\beta}d\tau_{i_1}\ldots d\tau_{i_l}d\tau_{j_1}\ldots d\tau_{j_l}\sum_{k_{i_1} p_{i_1} \alpha_{i_1}\beta_{i_1}}\ldots\sum_{k_{j_l} p_{j_l} \alpha_{j_l}\beta_{j_l}} e^{-i(\phi_{i_1}+\ldots+\phi_{i_l}-\phi_{j_1}-\ldots-\phi_{j_l})}\notag\\
 &\qquad\qquad\times\left\langle \left(\bar{c}_{p_{i_1}\beta_{i_1}}d_{k_{i_1}\alpha_{i_1}}\ldots\bar{c}_{p_{i_l}\beta_{i_l}}d_{k_{i_l}\alpha_{i_l}}\right)\left(\bar{d}_{k_{j_1}\alpha_{j_1}}c_{p_{j_1}\beta_{j_1}}\ldots\bar{d}_{k_{j_l}\alpha_{j_l}}c_{p_{j_l}\beta_{j_l}}\right)\right\rangle,
\end{align}
\end{widetext}
where $(i_1,\ldots,i_n)$ and $(j_1,\ldots,j_n)$ are permutations of $(1,\ldots,2l)$. Noting that the propagator of the electrons in the dot and electrons in the lead are independent of the channel index, {\it i.e} $\left\langle \bar{d}_{k_1\alpha_1}(\tau_1)d_{k_2\alpha_2}(\tau_2)\right\rangle=\delta_{\alpha_1\alpha_2}\left\langle \bar{d}_{k_1}(\tau_1)d_{k_2}(\tau_2)\right\rangle$ and $\left\langle \bar{c}_{p_1\beta_1}(\tau_1)c_{p_2\beta_2}(\tau_2)\right\rangle=\delta_{\beta_1\beta_2}\left\langle \bar{c}_{p_1}(\tau_1)c_{p_2}(\tau_2)\right\rangle$, and furthermore assuming a continuous spectrum ({\it i.e.} large lead and dot) 
such that $\sum_{k_1 k_2}\left\langle \bar{d}_{k_1}(\tau_1)d_{k_2}(\tau_2)\right\rangle$ and $\sum_{p_1 p_2}\left\langle \bar{c}_{p_1}(\tau_1)c_{p_2}(\tau_2)\right\rangle$ are proportional to each other, we obtain
\begin{widetext}
\begin{align}
 {\cal A}=
 &\ t^{2l}\left({\cal N M}\right)^l\int_{0}^{\beta}d\tau_1\ldots d\tau_{2l}e^{-i(\phi_1+\ldots+\phi_l-\phi_{l+1}-\ldots-\phi_{2l})}\sum_{p_1\ldots p_{2l}}\sum_{k_1\ldots k_{2l}}\notag\\
 &\qquad\qquad\times\left\langle \bar{c}_{p_1}\ldots\bar{c}_{p_l} c_{p_{l+1}}\ldots c_{p_{2l}}\right\rangle\left\langle d_{k_1}\ldots d_{k_l}\bar{d}_{k_{l+1}}\ldots\bar{d}_{k_{2l}}\right\rangle.
\end{align}
\end{widetext}
In the last line we use the fact that for every complete contraction of the above correlators, the number of free indices for $\alpha$ and $\beta$ reduces from $2l$ to $l$. Summing over each $\alpha$ we get a factor ${\cal N}$ and similarly summing over each $\beta$ we get a factor ${\cal M}$. Since there are $l$ such sums, we get an overall factor $\left({\cal N M}\right)^l$. 
 
 Thus, the partition function reduces to
 \begin{widetext}
 \begin{align}
  Z &=\sum_{l=0}^{\infty} \frac{\left({t^2\cal N M}\right)^l}{(l!)^2}\sum_{n=-\infty}^{\infty} e^{i2\pi n N_0}\int_{0}^{\beta}d\tau_1\ldots d\tau_{2l}\left[\int_{\phi(\beta)=\phi_0+2\pi n}[ {\cal D} \phi]e^{-\frac{1}{4 E_c}\int_{0}^{\beta}d\tau \left(\partial_{\tau}\phi(\tau)\right)^2-i(\phi_1+\ldots+\phi_l-\phi_{l+1}-\ldots-\phi_{2l})}\right]\notag\\
 &\qquad\times\sum_{p_1\ldots p_{2l}}\sum_{k_1\ldots k_{2l}}\left\langle \bar{c}_{p_1}\ldots\bar{c}_{p_l} c_{p_{l+1}}\ldots c_{p_{2l}}\right\rangle\left\langle d_{k_1}\ldots d_{k_l}\bar{d}_{k_{l+1}}\ldots\bar{d}_{k_{2l}}\right\rangle.
\label{eq:partition_intermediate}
 \end{align} 
 The next step is now to use a ``charge'' representation of the partition function.
 
We then transform the $\phi$ dependent part of the partition function (within square brackets) to the {\it charge} representation as follows
\begin{align}
 &\sum_{n=-\infty}^{\infty}e^{i2\pi n N_0}\int_{\phi(\beta)=\phi_0+2\pi n}[ {\cal D} \phi]e^{-\frac{1}{4 E_c}\int_{0}^{\beta}d\tau \left(\partial_{\tau}\phi(\tau)\right)^2-i(\phi_1+\ldots+\phi_l-\phi_{l+1}-\ldots-\phi_{2l})}\notag\\
 &=\sum_{n=-\infty}^{\infty}e^{i2\pi n N_0}\int[ {\cal D} Q]\int_{\phi(\beta)=\phi_0+2\pi n}[ {\cal D} \phi]\exp\left[-E_c\int_{0}^{\beta}d\tau \left\{Q^2+i\dot{Q}\phi\right\}+i 2\pi n Q(\beta)-i(\phi_1+\ldots+\phi_l-\phi_{l+1}-\ldots-\phi_{2l})\right],
 \label{eq:charge_intro}
\end{align}
where we have introduced the auxiliary bosonic field $Q(\tau)$, such that $Q(\beta)=Q(0)$. Defining the source term
\begin{align}
 J(\tau;\tau_1,\ldots,\tau_{2l})=\left[\delta(\tau-\tau_1)+\ldots+\delta(\tau-\tau_l)-\delta(\tau-\tau_{l+1})-\ldots-\delta(\tau-\tau_{2l})\right],
 \label{eq:source}
\end{align}
we can rewrite Eq.\ \eqref{eq:charge_intro} as 
\begin{align}
 &\int[ {\cal D} Q]\sum_{n=-\infty}^{\infty}e^{i 2\pi n \left[Q(0)+N_0\right]}e^{-E_c\int_{0}^{\beta}d\tau Q^2}\int_{\phi(\beta)=\phi_0+2\pi n}[ {\cal D} \phi]\exp\left[i\int_{0}^{\beta}d\tau\left\{\dot{Q}-J(\tau;\tau_1,\ldots,\tau_{2l})\right\}\phi\right]\notag\\
 &=\int[ {\cal D} Q]\sum_{n=-\infty}^{\infty}\delta\left(Q_0+N_0+n\right)e^{-E_C\int_{0}^{\beta}d\tau Q^2}\delta\left[\dot{Q}-J(\tau;\tau_1,\ldots,\tau_{2l})\right]\notag\\
 &=\sum_{n=-\infty}^{\infty}\int_{-N_0-n}^{-N_0-n}[ {\cal D} Q]e^{-E_c\int_{0}^{\beta}d\tau Q^2}\delta\left[Q(\tau)+N_0+n-\sum_{i=1}^{l}\Theta(\tau-\tau_i)+\sum_{i=l+1}^{2l}\Theta(\tau-\tau_i)\right]\notag\\
  &=\sum_{n=-\infty}^{\infty}\int_{n}^{n}[ {\cal D} Q]e^{-E_c\int_{0}^{\beta}d\tau \left(Q(\tau)-N_0\right)^2}\delta\left[Q(\tau)-n+\sum_{i=1}^{l}\Theta(\tau-\tau_i)-\sum_{i=l+1}^{2l}\Theta(\tau-\tau_i)\right],
\label{eq:blips}
  \end{align}
\end{widetext}
where in the last step we used the transformation $Q(\tau)\rightarrow N_0-Q(\tau)$. 

We now focus on the charge degeneracy point $N_0=1/2$ and assume that we are at very low temperatures ({\it i.e.} $\beta E_c\gg 1$ or more precisely $\beta \Delta^*>1$). This implies that we can focus only on the lowest two charge (energy) states. This mean trajectories of $Q(\tau)$ are restricted to access these two charge states $Q(\tau)=0,1$. 
This projection to the lowest energy states causes the time labels in Eq.\ \eqref{eq:partition_intermediate} to be nested, i.e. the trajectory of the charge (in imaginary time) consists of  a succession of blips. Let us focus on one such trajectory is given by the nesting $\tau_1>\tau_{l+1}>\tau_2>\tau_{l+2}>\ldots>\tau_{2l}$, and rename $\tau_{l+1}\rightarrow\tau'_1$,$\ldots$,$\tau_{2l}=\tau'_l$.  It is simple to see that permuting the time labels of the various nestings give equivalent contributions. There are $l!\times l!$ equivalent contributions, corresponding to permutations of the labels $\left(\tau_1,\ldots,\tau_l\right)$ and $\left(\tau'_1,\ldots,\tau'_l\right)$. Furthermore, the contribution from the trajectories $0\rightarrow 1\rightarrow0\rightarrow\ldots\rightarrow0$ and $1\rightarrow0\rightarrow 1\rightarrow\ldots\rightarrow 1$ are identical. Thus in this regime we get
\begin{widetext}
\begin{align}
   Z &=2e^{-\frac{\beta E_c}{4}}\sum_{l=0}^{\infty} \frac{\left({t^2\cal N M}\right)^l}{(l!)^2} l!\times l!\int_{0}^{\beta}d\tau_1\int_{0}^{\tau_1}d\tau'_1\ldots \int_{0}^{\tau'_{l-1}}d\tau_l\int_{0}^{\tau_l}d\tau'_l\notag\\
 &\qquad\times\sum_{p_1\ldots p_l p'_1\ldots p'_l }\sum_{k_1\ldots k_{l}k'_1\ldots k'_l}\left\langle \bar{c}_{p_1}(\tau_1)\ldots\bar{c}_{p_l}(\tau_l) c_{p'_{1}}(\tau_1)\ldots c_{p'_{l}}(\tau'_l)\right\rangle\left\langle d_{k_1}(\tau_1)\ldots d_{k_l}(\tau_l)\bar{d}_{k'_{1}}(\tau'_{1})\ldots\bar{d}_{k'_{l}}(\tau_{2l})\right\rangle.
\end{align}
We recall the identities
\begin{subequations}
\begin{align}
 \sum_{p_1 p_2}G_{p_1 p_2}(\tau_1-\tau_2)=\sum_{p_1 p_2}\left\langle c_{p_1}\bar{c}_{p_2}\right\rangle=\nu_0\lim_{\beta\rightarrow\infty}\frac{\pi/\beta}{\sin\left(\pi/\beta (\tau_1-\tau_2)\right)}\rightarrow\nu_0\frac{1}{\tau_1-\tau_2}
\end{align}
 and
 \begin{align}
  \sum_{k_1 k_2}G_{k_1 k_2}(\tau_1-\tau_2)=\sum_{k_1 k_2}\left\langle d_{p_1}\bar{d}_{p_2}\right\rangle=\nu_1\lim_{\beta\rightarrow\infty}\frac{\pi/\beta}{\sin\left(\pi/\beta (\tau_1-\tau_2)\right)}\rightarrow\nu_1\frac{1}{\tau_1-\tau_2},
 \end{align}
 \end{subequations}
where $\nu_0$ and $\nu_1$ denote the electron density of states in the lead and dot respectively. Thus we obtain
 \begin{align}
   Z &=2e^{-\frac{\beta E_c}{4}}\sum_{l=0}^{\infty} \left(\nu_0\nu_1 t^2\cal N M\right)^l\int_{0}^{\beta}d\tau_1\int_{0}^{\tau_1}d\tau'_1\ldots \int_{0}^{\tau'_{l-1}}d\tau_l\int_{0}^{\tau_l}d\tau'_l\left[\sum_{P}(-1)^P\frac{1}{\left(\tau'_1-\tau_{P_1}\right)\left(\tau'_2-\tau_{P_2}\right)\ldots\left(\tau'_l-\tau_{P_l}\right)}\right]^2,
\label{eq:almost}
   \end{align}
where $(P_1,\ldots,P_L)$ is a permutation of $(1,\ldots,l)$ and $(-1)^P$ is the corresponding sign (Note : interchanging a pair indices in $(1,\dots,l)$ gives a factor of $-1$). Neglecting the constant factor of $2e^{-\frac{\beta E_c}{4}}$ Eq.\ \eqref{eq:almost} reduces to
 \begin{align}
   Z &=\sum_{l=0}^{\infty} \left(\nu_0\nu_1 t^2\cal N M\right)^l\int_{0}^{\beta}d\tau_1\int_{0}^{\tau_1}d\tau'_1\ldots \int_{0}^{\tau'_{l-1}}d\tau_l\int_{0}^{\tau_l}d\tau'_l\left[\frac{\left[\prod_{i<j}\left(\tau_i-\tau_{j}\right)\right]\left[\prod_{m< n}\left(\tau'_m-\tau'_n\right)\right]}{\left[\prod_{i_1=1}^{l}\left(\tau'_1-\tau_{i_1}\right)\right]\ldots\left[\prod_{i_l=1}^{l}\left(\tau'_l-\tau_{i_l}\right)\right]}\right]^2.
\label{eq:semifinal}
   \end{align}

Renaming $\tau'_1\rightarrow\tau_2$, $\tau_2\rightarrow\tau_3$, $\ldots$, $\tau'_l\rightarrow\tau_{2l}$ and introducing an ultraviolet cutoff $1/\tau_c=\Delta^*$, we can rewrite this expression as
\begin{align}
    Z &=\sum_{l=0}^{\infty} \left(\nu_0\nu_1 t^2\cal N M\right)^l\int_{0}^{\beta}\frac{d\tau_1}{\tau_c}\int_{0}^{\tau_1-\tau_c}\frac{d\tau_2}{\tau_c}\ldots \int_{0}^{\tau_{2l-1}-\tau_c}\frac{d\tau_{2l}}{\tau_c}\exp\left[2\sum_{i<j}(-1)^{i+j}\log\left|\frac{\tau_i-\tau_j}{\tau_c}\right|\right].
    \label{eq:RC_Kondo}
\end{align}
Recalling the partition function of the completely anisotropic single-channel Kondo model (in the language of the Coulomb gas expansion)  
\begin{align}
  Z_{\text{cg}} &=\sum_{n=0}^{\infty} \left(\frac{J_{\perp}\nu}{2}\right)^{2n}\int_{0}^{\beta}\frac{d\tau_1}{\tau_c}\int_{0}^{\tau_1-\tau_c}\frac{d\tau_2}{\tau_c}\ldots \int_{0}^{\tau_{2n-1}-\tau_c}\frac{d\tau_{2n}}{\tau_c}\exp\left[2\sum_{i<j}(-1)^{i+j}\log\left|\frac{\tau_i-\tau_j}{\tau_c}\right|\right],
\end{align}
we show that it is identical to Eq.\ \eqref{eq:RC_Kondo}, where $\nu$ is the density of states of the electrons of the Kondo model. The relationship between the one-channel Kondo theory and the parameters of the quantum RC circuit can be easily deduced
\begin{align}
 \left(\frac{J_{\perp}\nu}{2}\right)^2=\nu_0\nu_1 t^2\cal N M.
\end{align}
\end{widetext}

\section{\label{app:C}Perturbation theory near perfect transmission}
It is convenient to introduce the Majorana fermions
\\
\\
\begin{align}
d_+&=d^{\dag}+d\notag\\
d_-&=i(d^{\dag}-d)\notag\\
\psi_+(x)&=\psi^{\dag}_s(x)+\psi_s(x)\notag\\
\psi_-(x)&=i\left[\psi^{\dag}_s(x)+\psi_s(x)\right],
\end{align}
which obey the commutation relations $\{\psi_{\zeta},\psi_{\zeta'}\}=2\delta_{\zeta\zeta'}$ and $\{d_{\zeta},d_{\zeta'}\}=2\delta_{\zeta\zeta'}$. Here, we have used the symbols $\zeta$ ($\zeta'$)=$\pm$. Using this representation
and integrating out the charge field away from $x=0$, the partition function of the system is given as a coherent state functional integral  
\begin{align}
Z=\int{\cal D}\hat{\phi}_c{\cal D}\psi_+{\cal D}\psi_-{\cal D}d_+{\cal D}d_-\exp[-S].
\end{align}
We rewrite the action $S=S_0+S_{\text{int}}$ using Eqs. \eqref{eq:action_origin}, \eqref{eq:spin_refermionized} and \eqref{eq:backscattering_refermionized} as
\begin{widetext}
\begin{align}
S_0&=S^{c}_0+\int_0^{\beta}\sum_{\zeta=\pm}\left[\int_{-\infty}^{\infty}dx\left(\psi_{\zeta}(x,\tau)\partial_\tau \psi_{\zeta}(x,\tau)-i \frac{v_F}{4}\psi_{\zeta}(x,\tau)\partial_x\ \psi_{\zeta}(x,\tau)\right)+\bar{d}_{\zeta}(\tau)\partial_\tau d(\tau)+iX_{\zeta}\psi_{\zeta}(0,\tau)d_{-\zeta}(\tau)\right].
\end{align} 
and
\begin{align}
S_{\text{int}}=\int_0^{\beta}d\tau\sum_{\zeta=\pm}\left(|\lambda_1|+\zeta |\lambda_2|\right)\hat{C}_{\zeta}(\tau)\psi_{\zeta}(0,\tau)d_{-\zeta}(\tau):=\int_0^{\beta} d\tau H_{\text{int}}(\tau).
\end{align}
We define the constants $\Gamma_\zeta=2 X_{\zeta}^2/v_F$, which describe the effective widths of the resonances of the (impurity) Majorana fermions. For the Majorana description near weak tansmission it is necessary to introduce the infrared cutoff $\Gamma=\max\{\Gamma_+,\Gamma_-\}$. This corresponds to the lifetime of the shorter lived Majorana fermion. The local ({\it i.e} at $x=0$) Green's function of interest for the noninteracting Hamiltonian can be computed from the equations of motion on the lines of  Ref. \onlinecite{LeHur}, and are given by
\begin{align}
G_{\zeta}(\omega_n)&=-\int_{0}^{\beta}d\tau e^{i\omega_n\tau}\left\langle T_{\tau}\left[\psi_{\zeta}(0,\tau)\psi_{\zeta}(0,0)\right]\right\rangle=-\frac{i}{v_F}\frac{\omega_n\sgn(\omega_n)}{\left(\omega_n+\Gamma\sgn(\omega_n)\right)}\\
C_{-\zeta}(\omega_n)&=-\int_{0}^{\beta}d\tau e^{i\omega_n\tau}\left\langle T_{\tau}\left[\psi_{-\zeta}(0,\tau)d_{\zeta}(0)\right]\right\rangle=-\frac{i2X_{-\zeta}}{v_F}\frac{\sgn(\omega_n)}{\left(\omega_n+ \Gamma\sgn(\omega_n)\right)}\\
D_{\zeta}(\omega_n)&=-\int_{0}^{\beta}d\tau e^{i\omega_n\tau}\left\langle T_{\tau}\left[d_{\zeta}(\tau)d_{\zeta}(0)\right]\right\rangle=-i\frac{2}{\omega_n+\Gamma\sgn(\omega_n)}.
\end{align}
Note, $\omega_n=(2n+1)\pi/\beta$ denote (fermionic) Matsubara frequencies.  Next, we compute the leading order correction to the charge response function
\begin{align}
K(\tau)=\left\langle T_{\tau}\left[N(\tau)N(0)\right]\right\rangle=\frac{2}{\pi^2}\left\langle T_{\tau}\left[\hat{\phi}^{l}_c(\tau)\hat{\phi}^{l}_c(0)\right]\right\rangle=K_0(\tau)+K_1(\tau)+K_2(\tau)+O(\lambda^3)
\end{align}
where we have defined the functions
\begin{align}
K_0(\tau)&=\frac{2}{\pi^2}\left\langle T_{\tau}\left[\hat{\phi}^{l}_c(\tau)\hat{\phi}^{l}_c(0)\right]\right\rangle_0\notag\\
K_1(\tau)&=-\frac{2}{\pi^2}\int_0^{\beta}ds_1\left\langle  T_{\tau}\left[\hat{\phi}^{l}_c(\tau)\hat{\phi}^{l}_c(0)H_{\text{int}}(s_1)\right]\right\rangle_0\notag\\
K_2(\tau)&=\frac{1}{\pi^2}\int_0^{\beta}ds_1 ds_2\left\langle T_{\tau}\left[\hat{\phi}^{l}_c(\tau)\hat{\phi}^{l}_c(0)H_{\text{int}}(s_1)H_{\text{int}}(s_2)\right]\right\rangle_0.
\end{align}
Here, the subscript $\langle\ldots\rangle_0$ implies that the expectation value is taken with respect to the quadratic action $S_0$. The contribution $K_0(\tau)$ was computed in Eq.\ \eqref{eq:K0_reference} from which we inferred $R_q=h/(2e^2)$. Let us explicitly denote the operators $\hat{C}_{\zeta}$ as $\hat{C}_{\zeta}=u_{\zeta}\hat{\phi}_{c}+v_{\zeta}\hat{\phi}^2_{c}$. The corrections to the Green's function due to backscattering from the function $K_1(\tau)$, is given by 
\begin{align}
K_1(i\omega_n)&=\frac{1}{\pi^2}\int_0^{\beta}ds e^{i\omega_n\tau}\sum_{\zeta=\pm}v_{\zeta}\left(|\lambda_1|+\zeta |\lambda_2|\right)\left\langle T_{\tau}\left[\hat{\phi}^{l}_c(\tau)\hat{\phi}^{l}_c(0)\hat{\phi}^l_c(s)^2\right]\right\rangle_0\left\langle T_{\tau}\left[\psi_{\zeta}(0,s_1) d_{-\zeta}(\zeta)(0)\right]\right\rangle_0\notag\\
&=\frac{\gamma}{2\pi E_c}\frac{1}{\left(1+\frac{\pi\omega_n}{2E_c}\right)^2} \left[\sum_{\zeta= \pm} \kappa_{\zeta}(N_0)^2\left(r_1 + \zeta r_2\right)^2\log\left(\frac{E_c}{\Gamma}\right)\right].
\end{align}
Here, we assume that we are in the Coulomb-blockaded regime and identify $E_c$ to be the ultraviolet cutoff . We define the functions $\kappa_+(N_0):=\cos(\pi N_0)$ and $\kappa_-(N_0):=\sin(\pi N_0)$. The energy scale of the Fermi velocity, which is lost in the bosonization process where an infinite bandwidth is assumed, is reintroduced by fixing the cutoff \cite{Flensberg}. The backscattering amplitudes are then determined to be
\begin{align}
\lambda_{1,2}=r_{1,2}\sqrt{\frac{ 2 a v_F E_C\gamma}{\pi}},
\end{align}
The contribution from the function $K_2(\tau)$ involves 3 types of diagrams
\begin{align}
K_2(i\omega_n)&=\frac{1}{\pi^2}  \int_0^\beta d\tau ds_1  ds_2 e^{i\omega_n\tau} \notag\\
    &\bigg[\sum_{\zeta = \pm} \left(\frac{\lambda_1 + \zeta \lambda_2}{\sqrt{2 \pi a}} \right)^2 \left\langle T_\tau [\hphi_c(\tau) \hphi_c(0)  \hat{C}_{\zeta}(s_1) \hat{C}_{\zeta}(s_2)]\right\rangle \left\langle T_\tau [\psi_{\zeta}(0,s_1)  \psi_{\zeta}(0,s_2)]\right\rangle \left\langle T_\tau [d_{-\zeta}(s_1) d_{-\zeta}(s_2)] \right\rangle \notag \\
&
    +\sum_{\zeta = \pm}\left(\frac{\lambda_1 + \zeta \lambda_2}{\sqrt{2 \pi a}} \right)^2\left\langle T_\tau[\hphi_c(\tau) \hphi_c(0)  \hat{C}_{\zeta}(s_1) \hat{C}_{\zeta}(s_2)]\right\rangle \left\langle T_\tau [\psi_{\zeta}(0,s_1)  d_{-\zeta}(s_2)]\right\rangle \left\langle T_\tau [\psi_{\zeta}(0,s_2)d_{-\zeta}(s_1)]\right\rangle \notag\\
   -\sum_{\zeta_{1,2} = \pm}&\left(\frac{\lambda_1 + \zeta_1 \lambda_2}{\sqrt{2 \pi a}} \right)\left(\frac{\lambda_1 + \zeta_2 \lambda_2}{\sqrt{2 \pi a}} \right)\left\langle T_\tau [\hphi_c(\tau) \hphi_c(0)  \hat{C}_{\zeta_1}(s_1) \hat{C}_{\zeta_2}(s_2)] \right\rangle \left\langle T_\tau [\psi_{\zeta_1}(0,s_1)  d_{-\zeta_1}(s_1)]\right\rangle \left\langle T_\tau [\psi_{\zeta_2}(0,s_2) d_{-\zeta_2}(s_2)] \right\rangle\bigg]
\end{align}
It is straightforward to show that the last term, which involves a product of cross-correlators of the impurity and lead Majorana fermions, is $O(\lambda^4)$ and can be neglected. Furthermore, in this regime we should neglecting correlators in $\hphi_c$ which involve more than a product of four $\hphi_c$ fields, since the charge is pinned due to strong Coulomb interactions. Thus we obtain
\begin{align}
K_2(i\omega_n)=-\frac{\gamma}{2\pi E_c}\frac{1}{\left(1+\frac{\pi\omega_n}{2E_c}\right)^2}\sum_{\zeta = \pm} \kappa_{-\zeta}(N_0)^2 \left(r_1 + \zeta r_2\right)^2\left[\log\left(\frac{E_c}{\Gamma}\right)-\left(1+\frac{2\Gamma}{\omega_n}\right)\log\left(1+\frac{\omega_n}{\Gamma}\right)\right].
\end{align}
In our calculations we make use of the Fourier transform of the $\hat{\phi}^{l}_c$ propagator 
\begin{align}
\tilde{F}_c(i\nu_k)=-\int_{0}^{\beta}d\tau e^{i\nu_k\tau}\left\langle T_{\tau}\left[\hat{\phi}^{l}_c(\tau)\hat{\phi}^{l}_c(0)\right]\right\rangle_0=-\frac{\pi}{2}\frac{1}{\left|\nu_k\right|+2E_C/\pi},
\end{align}
where $\nu_k=2k\pi/\beta$ denotes (bosonic) Matsubara frequencies. Wick rotating $i\omega_n\rightarrow \omega+i\delta$ we obtain
\begin{align}
K_1(\omega)&=\frac{\gamma}{2\pi E_c}\frac{1}{\left(1-i\frac{\pi\omega}{2E_c}\right)^2} \left[\sum_{\zeta= \pm} \kappa_{\zeta}(N_0)^2\left(r_1 + \zeta r_2\right)^2\log\left(\frac{E_c}{\Gamma}\right)\right]\notag\\
K_2(\omega)&=-\frac{\gamma}{2\pi E_c}\frac{1}{\left(1-i\frac{\pi\omega}{2E_c}\right)^2}\sum_{\zeta = \pm} \kappa_{-\zeta}(N_0)^2 \left(r_1 + \zeta r_2\right)^2\left[\log\left(\frac{E_c}{\Gamma}\right)-\left(1+i\frac{2\Gamma}{\omega}\right)\log\left(1-i\frac{\omega}{\Gamma}\right)\right].
\end{align}
\end{widetext}

\end{document}